\documentclass[11pt]{amsart}

\usepackage[T1]{fontenc}
\usepackage{amsmath,amssymb,amsthm,mathtools}
\usepackage{booktabs}
\usepackage{enumitem}
\usepackage[colorlinks=true,linkcolor=blue,citecolor=blue,urlcolor=blue]{hyperref}

\usepackage[margin=2cm]{geometry}

\newtheorem{theorem}{Theorem}[section]
\newtheorem{proposition}[theorem]{Proposition}
\newtheorem{lemma}[theorem]{Lemma}
\newtheorem{corollary}[theorem]{Corollary}
\theoremstyle{definition}
\newtheorem{definition}[theorem]{Definition}
\theoremstyle{remark}
\newtheorem{remark}[theorem]{Remark}
\newtheorem{example}[theorem]{Example}

\newcommand{\R}{\mathbb R}
\newcommand{\Sph}{\mathbb S}

\newcommand{\Fpos}{\mathcal F_{\ge 0}^{(n)}}

\title{Levenshtein-type bounds for $(a,b)$-avoiding spherical codes}

\author[P. G. Boyvalenkov]{P. G. Boyvalenkov} 
\address{ Institute of Mathematics and Informatics, Bulgarian Academy of Sciences \\
8 G Bonchev Str., 1113 Sofia, Bulgaria}
\email{peter@math.bas.bg}

\author[D. D. Cherkashin]{D. D. Cherkashin} 
\address{ Institute of Mathematics and Informatics, Bulgarian Academy of Sciences \\
8 G Bonchev Str., 1113 Sofia, Bulgaria}
\email{jiocb@math.bas.bg}

\author[P. D. Dragnev]{P. D. Dragnev}
\address{ Department of Mathematical Sciences, Purdue University \\
Fort Wayne, IN 46805, USA }
\email{dragnevp@purdue.edu}

\author[P. J. Grabner]{P. J. Grabner}
\address{ Institute of Analysis and Number Theory, Graz University of Technology, Kopernikusgasse 24. 8010 Graz, Austria}
\email{peter.grabner@tugraz.at}

\date{}

\begin{document}

\begin{abstract}
We develop a Levenshtein-type linear programming framework (for maximal $s$-codes and for minimal energy) for $(a,b)$-avoiding spherical codes, that is, codes whose inner products between distinct points avoid $(a,b)$. We introduce and investigate signed measures that are positive definite up to an appropriate degree and yield stationary auxiliary linear programming polynomials and exact quadrature formulas. We give explicit conditions for positivity of the quadrature weights and sufficient Krein-type conditions for nonnegative Gegenbauer coefficients (positive definiteness of the candidate optimal linear programming polynomials). Under these hypotheses and suitable node-location assumptions, the resulting cardinality bounds are optimal within the corresponding degree restriction and unique under natural nondegeneracy conditions. Under proper conditions on the quadrature nodes, we obtain universal lower energy bounds by utilizing Hermite interpolants as polynomial solutions of the respective linear program. Common test functions characterize optimality in the full polynomial linear programs under mild additional assumptions. 
\end{abstract}

\maketitle

\section{Introduction}

Bounds for the cardinality and energy of finite point sets on spheres are closely related problems in discrete geometry. The linear programming method of Delsarte, Goethals, and Seidel~\cite{DGS} (see also~\cite[Chapter 5]{BHS2019},~\cite[Chapter 9]{CS},~\cite[Chapter 2]{EZ2001},~\cite{KL78}) converts restrictions on pairwise inner products into inequalities for Gegenbauer expansions. Levenshtein's bounds~\cite{Lev79} (see also~\cite{Lev83,Lev92,Lev98}) organize this method into families of auxiliary polynomials associated with orthogonal polynomials and positive quadrature formulas. These quadratures explain both the quality of the bounds (cf.~\cite[Theorem 5.39]{Lev98}) and their optimality among auxiliary polynomials of specified degrees (cf.~\cite[Theorem 3.1]{BDB1996},~\cite[Theorem 5.47]{Lev98}). 

A parallel theory applies to the energy
\[
 E_h(C)=\sum_{\substack{x,y\in C, \ x\ne y}}h(\langle x,y\rangle)
\]
of a spherical code $C$, where $h:[-1,1) \to \mathbb{R}$ is a potential function. The energy linear programming method of Yudin~\cite{Y1992} and the theory of universal optimality by Cohn and Kumar~\cite{CK2007} connect polynomial minorants of the potential with geometric properties of extremal configurations. Boyvalenkov, Dragnev, Hardin, Saff, and Stoyanova~\cite{BDHSS2016Energy} used Levenshtein quadratures to obtain {\em universal lower energy bounds} (ULB) for absolutely monotone potentials (see also~\cite{BDHSS2021Lifted}). 
The universality is understood in the sense of Levenshtein~\cite{Lev98} (there is a suitable bound for every code), in the sense of Cohn--Kumar~\cite{CK2007} (the bound is valid for all absolutely monotone potentials), and in the fact, that the interpolation nodes and weights depend on the geometric parameters rather than on the potential, and the same Levenshtein quadrature controls both cardinality and energy.

These LP frameworks extend to polynomial metric spaces~\cite{Lev92,Lev98,BDHSS2019PM} and to weighted spherical codes and designs
\cite{BBDHSS2025}. Related quadrature methods also yield universal upper energy bounds~\cite{BDHSS2020Upper} and bounds for potentials
of weighted designs and energies of discrete measures~\cite{BBDHSS2026}. There are strong relations of spherical-code bounds to Euclidean sphere packing, see~\cite{KL78,CZ2014}. 

In this paper, we consider the maximal cardinality and minimum energy problems in classes of codes where additional restrictions on inner products are imposed. If a set $T \subset [-1,1)$ is forbidden for inner products of a spherical code $C$, then we call $C$ a $T$-avoiding code. Delsarte, Goethals, and Seidel~\cite{DGS} considered $A$-codes which are codes with inner products in the set $A$; in our terminology, they would be $([-1,1) \setminus A)$-avoiding codes. The linear programming framework, developed in~\cite{DGS}, is then directly applied (see Theorem~\ref{LP-theorem} below). The $T$-avoiding setting is also related to a corresponding sphere packing setting, see~\cite{GV2025}. 

Boyvalenkov, Dragnev, Hardin, Saff, and Stoyanova~\cite{BDHSS2019Interval} obtained bounds for spherical codes with inner products in a prescribed interval $[\ell,s]$, i.e., for $[-1,\ell)$-avoiding $s$-codes. Their approach uses signed measures with positive definiteness up to appropriate degrees (a technique, already introduced in the field by Cohn and Kumar~\cite{CK2007}), associated orthogonal polynomials, and Krein conditions. More recently, Boyvalenkov, Cherkashin, and Dragnev~\cite{BCD2026} studied $T$-avoiding spherical codes and designs and established maximality of certain $T$-avoiding $s$-codes, minimality of certain $T$-avoiding spherical designs, and corresponding restricted universal optimality. These results motivate a systematic treatment of an excluded interval $T=(a,b)$, for which the allowed inner-product set has two components (in other words, interval-avoiding $s$-codes). 


Fix
\[
 n\ge2,\qquad -1\le a<b<s<1,
 \qquad X_{a,b,s}=[-1,a]\cup[b,s].
\]
We study the maximum cardinality $\mathcal A(n,s;a,b)$ of spherical codes whose inner products between distinct points belong to $X_{a,b,s}$. Thus the usual separation restriction $\langle x,y\rangle\le s$ for $x \neq y \in C$ is supplemented by the exclusion of the interval $(a,b)$. For energy, we consider $(a,b)$-avoiding codes of fixed cardinality without imposing an upper cutoff $s$ on their inner products. We write $\mathcal E_h(n,N;a,b)$ for the infimum of the energy over such codes of cardinality $N$. Their allowed set is $[-1,1)\setminus(a,b)$; the parameter $s$ instead enters through the quadrature used to construct the bound.

We adapt the signed measure and quadrature approach to this disconnected allowed set. The natural linear programming candidate polynomials have the forms
\[
 f(t)=(t-a)(t-b)(t-s)A_k(t)^2
\]
of degree $2k+3$, and
\[
 f^+(t)=(1+t)(t-a)(t-b)(t-s)B_k(t)^2
\]
of degree $2k+4$. Clearly, these polynomials are non-positive on $X_{a,b,s}$. The main task is to choose the square factors so that the Gegenbauer coefficients are nonnegative and the resulting bound is optimal. The two boundary points of the excluded interval introduce the signed measures
\[
 d\nu_{a,b}(t)=(1-t)(t-a)(t-b)\,d\mu_n(t),
 \qquad
 d\nu_{a,b}^{+}(t)=(1+t)\,d\nu_{a,b}(t),
\]
where $\mu_n$ is the normalized Gegenbauer measure.

Under the assumption of positive definiteness of $\nu_{a,b}$ ($\nu_{a,b}^+$) up to appropriate degree and the necessary normalization assumptions, we derive Euler equations and Christoffel--Darboux representations for the stationary factors $A_k$ and $B_k$. They generate quadrature formulas which are exact up to degrees $2k+3$ and $2k+4$, respectively. Exactness alone does not ensure that the nodes lie in the allowed set $X_{a,b,s}$ or that the quadrature weights are positive. We give explicit formulas for the weights and impose separate node-location and weight-positivity conditions. A column form of the necessary variant of the Krein condition then supplies sufficient conditions for the stationary polynomial to be LP-admissible. If $\rho_1$ denotes the quadrature weight at $1$, the cardinality bound can be written as 
\[
 \mathcal A(n,s;a,b)\le \frac{1}{\rho_1},
\]
with the analogous expression $1/\rho_1^+$ in the even branch (cf.~\cite[Section 5.4]{Lev98} for Levenshtein's analogs). Positive quadrature at admissible nodes proves optimality within the corresponding degree restriction; additional interior-node assumptions give uniqueness up to positive scaling.

In the energy setting, the natural linear programming candidate polynomials are Hermite interpolants to the potential (cf.~\cite{BDHSS2016Energy}). 
We derive sufficient conditions for a Newton ordering in which every required partial product in the Newton formula is positive definite. These conditions yield admissible energy minorants. If the quadrature parameter $N=1/\rho_1$ is an integer, we derive the universal lower bound (ULB) 
\[
 \mathcal E_h(n,N;a,b)\ge
 N^2\sum_{\xi\ne1}w_\xi h(\xi),
\]
where the sum runs over the nontrivial quadrature nodes $\xi$ and $w_\xi$ are their (positive) weights. The bound is universal in the same three senses as explained above as, more precisely said, the same nodes and weights work for every absolutely monotone potential. The applicability of the ULB remains conditional on the geometric and positivity hypotheses.

Finally, we study quadrature test functions in the spirit of~\cite{BDB1996,BDHSS2016Energy} (also~\cite[Theorem 5.47]{Lev98}). Their nonnegativity in all degrees beyond the quadrature precision certifies global optimality, i.e., optimality in the full polynomial LP. Like in the classical case, a negative value of a test function implies a higher-degree improvement when the optimal cardinality polynomial $f$ or $f^+$ has strictly positive Gegenbauer coefficients (which is the case with the Levenshtein polynomials in the classical case), or when the energy potential is strictly absolutely monotone. This distinguishes optimality at a fixed degree from optimality among all polynomial certificates, a distinction also central to higher-level extensions of the Levenshtein and classical ULB frameworks~\cite{BDHSS2021Lifted}.

The results imply a conditional framework (see the Appendix) rather than a classification of all admissible parameters. After setting up the linear programs, we develop the signed measures, stationary polynomials, quadrature formulas, and Krein criteria. We then turn to energy interpolation and test functions, and conclude with the examples and the remaining positivity questions. We illustrate the construction with two $23$-dimensional odd-degree
examples recovering known bounds from~\cite{BCD2026} and a degenerate $22$-dimensional even-degree example (again from~\cite{BCD2026}). For
each, we give the exact quadrature, the distance distribution of an
attaining design, and the corresponding sharp energy bound.
In the last case, a quadrature weight vanishes and the bound coincides with that of the neighboring odd degree branch (see Subsection~\ref{sec2025}). This phenomenon is not unexpected --- in the classical case it was proved by Boyvalenkov, Danev and Landjev~\cite{BDL1999} that the even degree Levenshtein bounds can be possibly attained only in the endpoints of their optimality and in that points they can be identified as odd bounds as well. 

The paper is organized as follows. In Section 2 we explain the necessary linear programming setting for the maximal codes problem. The necessary signed measures and their polynomials (orthogonal up to appropriate degrees) are defined in Section 3. Stationary polynomials for the maximal codes problem are found in Section 4. In Section 5, the relating quadrature formula is derived and investigated. In Section 6 we first introduce the (odd and even) analogs of the Krein condition used by Levenshtein in his derivation and then prove the Levenshtein-type bound on maximal $(a,b)$-avoiding codes. Section 7 is devoted to universal lower bounds on energy of codes as we first give the LP energy setting and then provide the corresponding bound; here a positive Newton ordering is needed to establish the bound. Section 8 is devoted to test functions which give a necessary and sufficient condition for global optimality of our bounds (both for maximal codes and minimum energy). Some examples are presented in Section 9 and an appendix gives a programme for implementation of numerics in our framework.

\section{Linear programming setting}

Let $n \geq 2$ and 
\[
 d\mu_n(t)=\gamma_n(1-t^2)^{(n-3)/2}\,dt, \ \ -1\le t\le 1,
\]
where $\gamma_n$ is chosen to make $\mu_n$ a probability measure; i.e., so that $\mu_n([-1,1])=1$. Let
$\left(P_i^{(n)}\right)_{i=0}^{\infty}$ denote the normalized Gegenbauer polynomials, orthogonal with respect to $\mu_n$, 
\[ \int_{-1}^{1}P_i^{(n)}(t)P_j^{(n)}(t)\,d\mu_n(t) =\frac{\delta_{ij}}{r_i}, \]
normalized by $P_i^{(n)}(1)=1$, where 
\[ r_i=\dim \operatorname{Harm}_i(\R^n) =\binom{n+i-1}{i}-\binom{n+i-3}{i-2}. \]
For the classical theory of Gegenbauer and Jacobi polynomials, see~\cite{Szego1975}; for recurrence relations and quadrature methods,
see also~\cite{Gautschi2004}.

Every real polynomial $f \in \mathbb{R}[t]$ is expanded uniquely\footnote{The corresponding infinite Gegenbauer expansion is actually true for $f \in L^2(\mu_n)$.} in terms of Gegenbauer polynomials 
\begin{equation} \label{Geg_exp}
 f(t)=\sum_{i=0}^{\deg f}f_iP_i^{(n)}(t).
\end{equation}
We are interested in positive definite polynomials. These are polynomials $f$ with nonnegative Gegenbauer coefficients in~\eqref{Geg_exp}. Denote the class of positive definite polynomials by
\[
 \Fpos
 :=\left\{f:\ f_i\ge0\text{ for every }i\right\}.
\]
By Schoenberg's characterization~\cite{S1942}, nonnegative Gegenbauer coefficients are equivalent to positive semidefiniteness of the zonal kernel $(x,y)\mapsto f(\langle x,y\rangle)$ on $\mathbb S^{n-1}$.

Good spherical codes are often spherical designs of high strength. Spherical designs were introduced by Delsarte, Goethals, and Seidel in the seminal paper~\cite{DGS}. A spherical code $C\subset \mathbb{S}^{n-1}$ is a {\em spherical $\tau$-design} if the average (arithmetic mean) over $C$ of any polynomial in $n$ variables of total degree at most $\tau$ equals the average with respect to the surface measure $\nu$ over the unit sphere (cf.~\cite[Section 5]{DGS}), i.e.,
\[ \frac{1}{|C|} \sum_{x \in C} f(x) = \frac{1}{\nu(\mathbb{S}^{n-1})}\int_{\mathbb{S}^{n-1}} f(x) \, d \, \nu(x) \]
for every $n$-variable polynomial $f$ of total degree at most $\tau$. Equivalently, and very conveniently for us, $C$ is a spherical $\tau$-design if and only if the {\em moments}
\[
M_i (C):=\sum_{x,y\in C}P_i^{(n)} (\langle x,y\rangle)
\]
vanish for all $i=1,\dots,\tau$. For surveys on spherical designs see~\cite{BB2009,AZ2027}.

We are interested in spherical codes and designs such that the inner products of their distinct points avoid some subsets of $[-1,1)$. 

\begin{definition}
For fixed $-1\le a<b<s<1$, denote by $X_{a,b,s}:=[-1,a]\cup[b,s]$ the set of \emph{allowed} inner products. A spherical code $C\subset\Sph^{n-1}$ is an $(a,b)$-\emph{avoiding} $s$-\emph{code} if
\[ \langle x,y\rangle\in X_{a,b,s} \quad \mbox{for every} \quad x,y\in C,\ x\ne y. \]
If the maximal inner product is not important, we speak about just $(a,b)$-avoiding codes, meaning that $\langle x,y\rangle \not\in (a,b)$ only. 
\end{definition}

Denote by 
\[ \mathcal A(n,s;a,b):= \max\{|C|: C \subset \mathbb{S}^{n-1} \text{ is an $(a,b)$-avoiding $s$-code}\} \]
the maximal cardinality of an $(a,b)$-avoiding $s$-code. 

We write
\[
 I(C):=\{\langle x,y\rangle:x,y\in C,\ x\ne y\}=\{t_1<\cdots<t_d\}
\]
for the set of inner products between distinct points of $C$. For $x\in C$ and $t\in[-1,1)$, put $A_t(x;C):=\bigl|\{y\in C\setminus\{x\}:\langle x,y\rangle=t\}\bigr|$.
The vector
\[
 F_x(C):=\bigl(A_{t_1}(x;C),\ldots,A_{t_d}(x;C)\bigr)
\]
is the distance distribution of $C$ with respect to $x$.
If it is independent of $x$, the code is called \emph{distance invariant}
(or \emph{regular}), and we write $A_t(C)$ and $F(C)$.
By~\cite[Theorem~7.4]{DGS}, a spherical $\tau$-design is distance invariant
whenever $\tau\ge d-1$. Indeed, for each $x\in C$, the design property gives
\begin{equation}\label{Vand}
 \sum_{i=1}^d A_{t_i}(x;C)t_i^r
 =|C|\int_{-1}^1t^r\,d\mu_n(t)-1,
 \qquad 0\le r\le d-1.
\end{equation}
 
This is a nonsingular {\em Vandermonde system} (the coefficient matrix is a Vandermonde matrix) whose right-hand side is
independent of $x$.

The linear programming variant for estimating the quantity $\mathcal A(n,s;a,b)$ was first derived in the same Delsarte--Goethals--Seidel paper~\cite{DGS} (see also~\cite{KL78,Lev79}).

\begin{theorem}[LP bound~\cite{DGS}] \label{LP-theorem}
Suppose that $f \in \mathbb{R}[t]$ is such that 

{\rm (A1)} $f(t)\le0$ for every $t\in X_{a,b,s}$,

{\rm (A2)} $f_0>0$, $f_i\ge0$ for every $i\ge1$.

\noindent Then
\[
 \mathcal A(n,s;a,b)\le\frac{f(1)}{f_0}.
\]
If $C \subset \mathbb{S}^{n-1}$ is an $(a,b)$-avoiding $s$-code, $f$ is a polynomial satisfying (A1) and (A2) and 
\[ |C|=\frac{f(1)}{f_0}=\mathcal A(n,s;a,b), \]
then $f_iM_i(C)=0$ for every $i \geq 1$ and $f(t)=0$ for every $t \in I(C)$. If $f_i>0$ for each $i=1,\ldots,\deg f$, then $C$ is a regular spherical $(\deg f)$-design.
\end{theorem}

We denote by $\mathcal{LP}_{n,s}^{a,b}$ the cone of polynomials satisfying the conditions (A1) and (A2); i.e.,
\[
 \mathcal{LP}_{n,s}^{a,b}:=\left\{ f(t)=\sum_{r=0}^{\deg f}f_rP_r^{(n)}(t): f(t)\le 0\ \text{for }t\in X_{a,b,s},\ f_0>0, \ f_r \ge0\ \text{for } r\ge1 \right\}.
\]

Following Levenshtein~\cite{Lev78,Lev79}, we search for optimal, in a sense, polynomials 
\begin{equation} \label{f-poly}
 f(t)= \left\{\begin{array}{ll}
 & (t-a)(t-b)(t-s)A_k^2(t), \mbox{ if $\deg(f)=2k+3$ is odd;} \\[4pt] 
 & (t+1)(t-a)(t-b)(t-s)B_k^2(t), \mbox{ if $\deg(f)=2k+4$ is even,} 
\end{array}\right.
\end{equation} 
in the set $\mathcal{LP}_{n,s}^{a,b}$. Here $A_k$ and $B_k$ are real polynomials of degree $k$ which will be optimized to give the best ratio $f(1)/f_0$ subject to $f \in \mathcal{LP}_{n,s}^{a,b}$. In fact, the condition (A1) is automatically satisfied by the polynomials~\eqref{f-poly} and we need to optimize $A_k$ and $B_k$ in such a way that $f$ satisfies (A2) and minimizes $f(1)/f_0$ (possibly up to the degree of $f$). 

In what follows we consider in detail the odd case $\deg f =2k+3$ and only record (sometimes with comments) the counterpart results for the even case using the superscript $+$ when necessary. In the even case we additionally assume that $a>-1$. 

\section{The signed measures \texorpdfstring{$d\nu_{a,b}$}{nu a,b} and \texorpdfstring{$d\nu_{a,b}^{+}(t):=(1+t)\,d\nu_{a,b}(t)$}{d nu+ a,b (t) = (1+t)dnu{a,b}(t)}}

We consider the measure 
\[
 d\nu_{a,b}(t):=(1-t)(t-a)(t-b)\,d\mu_n(t)
\]
and its even counterpart $d\nu_{a,b}^{+}(t):=(1+t)\,d\nu_{a,b}(t)=(1-t^2)(t-a)(t-b)\,d\mu_n(t)$. Both measures are signed because $(t-a)(t-b)<0$ on $(a,b)$. We use these signed measures under assumptions that they produce a sequence of polynomials which are orthogonal up to a certain degree. Denote by $\Pi_m$ the space of real one-variable polynomials of degree at most $m$ with the convention $\Pi_{-1}=\{0\}$. 

\begin{definition} 
A signed Borel measure $\mu$ on $\mathbb{R}$ for which all polynomials are integrable
is called {\it positive definite up to degree $m$} if for all real polynomials $p \not\equiv 0$
of degree at most $m$ we have $\int p^2 (t) d \mu(t) > 0$. For such $\mu$, the bilinear form
\begin{equation*}\label{ipsm}
\langle f, g\rangle_{\mu}:=\int f(t)g(t)\, d\mu(t),
\end{equation*}
is an inner product on the space $\Pi_m$.
\end{definition} 

Note that the positive definiteness up to degree $k$ of $\nu_{a,b}$ 
is equivalent to the positive definiteness of the Gram matrix
\[ G_k(a,b;n) :=\left( \int_{-1}^{1}t^{i+j}(1-t)(t-a)(t-b) \,d\mu_n(t) \right)_{i,j=0}^{k}.
\]

\begin{example} For fixed $n$, the Gram matrix condition leads to polynomial inequalities in $a$ and $b$. For $k=0,1,2$, the Sylvester criterion (a real symmetric matrix is positive definite if and only if all its leading principal minors are positive) implies that the measure $\nu_{a,b}$ is positive definite up to degree $i$, $i=0,1,2$, if and only if $\Delta_j(a,b;n)>0$ for $0 \leq j \leq i$, respectively. Here 
\[ \Delta_0(a,b;n)= \det G_0(a,b,n), \]
\[ \Delta_1(a,b;n)=\frac{n^2(n+2)^2}{n-1} \cdot \det G_1(a,b;n),
\] 
\[ \Delta_2(a,b;n)=
 \frac{n^3(n+2)^3(n+4)^2}{2(n-1)^2} \cdot \det G_2(a,b;n). \]
\end{example}

Assuming that $\nu_{a,b}$ is positive definite up to degree $k$, we apply the Gram--Schmidt orthogonalization procedure to construct a sequence of monic polynomials which we assume to be non-vanishing at $1$. Then we construct the unique sequence 
\[ R_0^{a,b},R_1^{a,b},\ldots,R_{k+1}^{a,b} \] 
orthogonal with respect to $\nu_{a,b}$ with normalization $R_i^{a,b}(1)=1$, $i=0,1,\ldots,k+1$. 
We also define normalization constants $r_i^{a,b}$ by 
\[ r_i^{a,b} \int_{-1}^{1} \left(R_i^{a,b}(t)\right)^2\,d\nu_{a,b}(t)=1, \ \ 0\le i\le k. \]
Then the Christoffel--Darboux kernel for our new sequence is
\begin{equation}\label{eq:CDkernel}
 T_k^{a,b}(t,u)
 :=\sum_{i=0}^{k} r_i^{a,b} R_i^{a,b}(t)R_i^{a,b}(u).
\end{equation}

The even case is dealt similarly. The corresponding Gram matrix is 
\begin{equation*}\label{eq:evenGram}
 G_k^+(a,b;n)
 :=\left(
 \int_{-1}^{1}t^{i+j}(1-t^2)(t-a)(t-b)\,d\mu_n(t)
 \right)_{i,j=0}^{k}.
\end{equation*}
Then the sequence $(R_i^{a,b,+}(t))_{i=0}^{k+1}$ with normalization $R_i^{a,b,+}(1)=1$, the constants $(r_i^{a,b,+})_{i=0}^k$, and the Christoffel--Darboux kernel $T_k^{a,b,+}(t,u)$ are defined analogously. 

We shall often assume that the measure $\nu_{a,b}$ ($\nu_{a,b}^+$, respectively) is positive definite up to the required degree rather than proving this. Of course, the positive definiteness can be readily checked via the above Gram matrix for any particular values of $n$, $a$ and $b$. 

\section{The stationary polynomials}

We find representations of the candidate optimal polynomials for Theorem~\ref{LP-theorem}. For a fixed degree bound, the full LP is convex after the normalization $f_0=1$. Below we derive stationary candidates in the factorized family;
LP-admissibility and degree-restricted optimality are established separately in Section~\ref{KreinSec}.

Recall that we describe in detail the odd case only. Fix a non-negative integer $k$ and consider
\begin{equation}\label{eq:Fdef}
 f(t)=(t-a)(t-b)(t-s)A_k^2(t), \ \deg A_k = k, \ A_k(1)=1.
\end{equation}
Then the numerator 
\[ f(1)=(1-a)(1-b)(1-s) \]
does not depend on the polynomial $A_k$. Thus minimizing $f(1)/f_0$ within the class of polynomials~\eqref{eq:Fdef} is equivalent to maximizing
\[ f_0 = \int_{-1}^{1} (t-a)(t-b)(t-s)A_k^2(t)\,d\mu_n(t). \]

\begin{proposition}[Euler equation, see, e.g., Chapter 1.4 in~\cite{GelfandFomin1963}] \label{prop:Euler}
A stationary point of the minimization problem satisfies
\begin{equation}\label{eq:Euler}
 \int_{-1}^{1}(t-s)A_k(t)U(t)\,d\nu_{a,b}(t)=0 \quad \mbox{for every} \quad U \in\Pi_{k-1}.
\end{equation}
Consequently, we have the representation
\begin{equation}\label{eq:quasiorthogonal}
 (t-s)A_k(t)
 =\gamma_s\bigl(R_{k+1}^{a,b}(t)+c_sR_k^{a,b}(t)\bigr),
\end{equation}
where the constants $c_s$ and $\gamma_s$ are given by
\begin{equation}\label{eq:cs}
 c_s=-\frac{R_{k+1}^{a,b}(s)}{R_k^{a,b}(s)}, \ \ 
 \gamma_s=\frac{1-s}{1+c_s},
\end{equation}
under the assumptions $R_k^{a,b}(s) \neq 0$ and $c_s>-1$. 
\end{proposition}

\begin{proof}
Every variation preserving the normalization condition $A_k(1)=1$ is of the form
\[ 
A_k(t)+\varepsilon(1-t)U(t), \ U\in\Pi_{k-1}.
\]
Differentiating $f_0 = f_0(\varepsilon)$ at $\varepsilon=0$ gives
\[ 0=2\int_{-1}^{1}(t-a)(t-b)(t-s)A_k(t)(1-t)U(t)\,d\mu_n(t), \]
which is~\eqref{eq:Euler}. Now the polynomial $(t-s)A_k(t)$ has degree $k+1$ and is orthogonal to
$\Pi_{k-1}$ with respect to the measure $\nu_{a,b}$. Therefore, it is a linear combination of $R_{k+1}^{a,b}$ and $R_k^{a,b}$, which gives
\eqref{eq:quasiorthogonal} up to the calculation of the constants. Evaluation at $t=s$ gives the formula for $c_s$;
evaluation at $t=1$, together with the normalizations $A_k(1)=R_k^{a,b}(1)=R_{k+1}^{a,b}(1)=1$, gives the
formula for $\gamma_s$ in~\eqref{eq:cs}.
\end{proof}

\begin{proposition}[Kernel representation]\label{prop:kernel}
Assume $R_k^{a,b}(s)\ne0$ and $T_k^{a,b}(1,s)\ne0$. Then the factor $A_k$ of normalized stationary polynomial of degree $2k+3$ is given by 
\begin{equation}\label{eq:AkKernel}
 A_k(t)=\frac{T_k^{a,b}(t,s)}{T_k^{a,b}(1,s)}.
\end{equation}
\end{proposition}

\begin{proof}
The Christoffel--Darboux identity has the form
\[
T_k^{a,b}(t,s) =d_k \frac{R_{k+1}^{a,b}(t)R_k^{a,b}(s)-R_k^{a,b}(t)R_{k+1}^{a,b}(s)}{t-s}
\]
with $d_k\ne0$. By~\eqref{eq:cs}, the numerator is proportional to
$R_{k+1}^{a,b}(t)+c_sR_k^{a,b}(t)$. Equations~\eqref{eq:quasiorthogonal} and
$A_k(1)=1$ now yield~\eqref{eq:AkKernel}.
\end{proof}

\begin{remark}
Under the positive definiteness up to degree $k$ assumption, the associated
finite Jacobi matrix, formed using the monic recurrence,
has real diagonal entries and positive off-diagonal entries.
For every real $c$, the polynomial
$R_{k+1}^{a,b}+cR_k^{a,b}$ is proportional to the
characteristic polynomial of such a matrix with a rank-one
change in its last diagonal entry. Hence it has $k+1$
real simple zeros. Positive definiteness up to degree $k$ alone does
\emph{not} force these zeros to belong to $[-1,1]$ or to avoid $(a,b)$; this
location of zeros must be established separately.
\end{remark}

In the even case, we obtain analogously the representations 
\begin{equation}\label{eq:evenQuasiorthogonal}
 (t-s)B_k(t)
 =\gamma_s^+
 \left(R_{k+1}^{a,b,+}(t)+c_s^+R_k^{a,b,+}(t)\right),
\end{equation}
where 
\begin{equation*}\label{eq:evenCsGamma}
 c_s^+
 =-\frac{R_{k+1}^{a,b,+}(s)}{R_k^{a,b,+}(s)},
 \qquad
 \gamma_s^+=\frac{1-s}{1+c_s^+},
\end{equation*}
and 
\[
 B_k(t)=\frac{T_k^{a,b,+}(t,s)}{T_k^{a,b,+}(1,s)},
\]
where $R_k^{a,b,+}(s)\neq 0$ and $T_k^{a,b,+}(1,s)\neq 0$.

\section{The two-point quadrature formula and its weights}

\subsection{The exact quadrature formula}

The roots of the polynomial $(t-s)A_k(t)$ along with $1$, $a$, and $b$ as fixed nodes define a key quadrature formula as follows. 
Assume that $s$ is the largest root of $(t-s)A_k(t)$.
Enumerate the roots as $\alpha_1<\cdots<\alpha_{k+1}=s$,
and assume that they are distinct from $1,a,b$.

\begin{theorem}[Exact two-point quadrature]\label{thm:quadrature}
There exist unique real weights $\rho_1,\rho_a,\rho_b,\lambda_1,\ldots,\lambda_{k+1}$
such that the quadrature formula 
\begin{equation}\label{eq:quadrature}
 p_0=\int_{-1}^{1}p(t)\,d\mu_n(t) =\rho_1 p(1)+\rho_a p(a)+\rho_b p(b)
 +\sum_{i=1}^{k+1}\lambda_i p(\alpha_i)
\end{equation}
is exact for every $p\in\Pi_{2k+3}$.
\end{theorem}

\begin{proof}
Let $p \in \Pi_{2k+3}$ be fixed and $I[p]$ be the Lagrange interpolation polynomial of degree at most $k+3$ that agrees
with $p$ at the $k+4$ nodes $1,a,b,\alpha_1,\ldots,\alpha_{k+1}$.
Then 
\[ p(t)-I[p](t)=(t-1)(t-a)(t-b)(t-s)A_{k}(t)q(t) \]
for some polynomial $q\in\Pi_{k-1}$. Hence, by~\eqref{eq:Euler},
\begin{equation} \label{eq:remainder}
\int_{-1}^{1}(p-I[p])\,d\mu_n
 =-\int_{-1}^{1}(t-s)A_{k}(t)q(t)\,d\nu_{a,b}(t) =0. 
\end{equation}
Integrating the Lagrange formula for $I[p]$ gives~\eqref{eq:quadrature}.
The uniqueness of the Lagrange interpolation implies the uniqueness of the weights.
\end{proof}

The weights in~\eqref{eq:quadrature} can be written as
\begin{align}
 \rho_1
 &=\int_{-1}^{1}\frac{(t-a)(t-b)(t-s)A_{k}(t)}{(1-a)(1-b)(1-s)}\,d\mu_n(t),\\
 \rho_a
 &=\int_{-1}^{1}
 \frac{(t-1)(t-b)(t-s)A_{k}(t)}
 {(a-1)(a-b)(a-s)A_{k}(a)}\,d\mu_n(t),\\
 \rho_b
 &=\int_{-1}^{1}
 \frac{(t-1)(t-a)(t-s)A_{k}(t)}
 {(b-1)(b-a)(b-s)A_{k}(b)}\,d\mu_n(t),\\
 \lambda_i
 &=\int_{-1}^{1}
 \frac{(t-1)(t-a)(t-b)(t-s)A_{k}(t)}
 {(t-\alpha_i)(\alpha_i-1)(\alpha_i-a)(\alpha_i-b)D_{i}(\alpha_i)}
 \,d\mu_n(t), \ \ i=1,\ldots,k+1,
\end{align}
where $D_i(t)$ is the derivative of $(t-s)A_k(t)$ (the apparent singularity at $\alpha_i$ in the formula for $\lambda_i$ is removable). These formulas are exact but do not by themselves make the signs transparent. We proceed below with the investigation of the positivity of the weights. 

In the even case, we assume that $s$ is the largest root of $(t-s)B_k(t)$ and denote all the roots of $(t-s)B_k(t)$ by
$ \beta_1<\beta_2<\cdots<\beta_{k+1}=s$, assuming that they are distinct from $-1,1,a,b$. Then there exist unique real weights
$\rho_{-1}^+,\ \rho_1^+,\ \rho_a^+,\ \rho_b^+, \ \lambda_1^+,\ldots,\lambda_{k+1}^+$ such that
\begin{equation} \label{eq:evenQuadrature}
 p_0=\int_{-1}^{1}p(t)\,d\mu_n(t)
 =\rho_{-1}^+p(-1)+\rho_1^+p(1)
 +\rho_a^+p(a)+\rho_b^+p(b)+\sum_{i=1}^{k+1}\lambda_i^+p(\beta_i) 
\end{equation}
for every $p\in\Pi_{2k+4}$. The Lagrange formulas for the weights are
\begin{align}
 \rho_1^+
 &=\int_{-1}^{1}
 \frac{(t+1)(t-a)(t-b)(t-s)B_k(t)}{2(1-a)(1-b)(1-s)}\,d\mu_n(t),
 \label{eq:evenRhoOneLagrange}\\
 \rho_{-1}^+
 &=\int_{-1}^{1}
 \frac{(t-1)(t-a)(t-b)(t-s)B_k(t)}
 {2(1+a)(1+b)(1+s)B_k(-1)}\,d\mu_n(t),
 \label{eq:evenRhoMinusLagrange}\\
 \rho_a^+
 &=\int_{-1}^{1}
 \frac{(t+1)(t-1)(t-b)(t-s)B_k(t)}
 {(a^2-1)(a-b)(a-s)B_k(a)}\,d\mu_n(t),
 \label{eq:evenRhoALagrange}\\
 \rho_b^+
 &=\int_{-1}^{1}
 \frac{(t+1)(t-1)(t-a)(t-s)B_k(t)}
 {(b^2-1)(b-a)(b-s)B_k(b)}\,d\mu_n(t),
 \label{eq:evenRhoBLagrange}\\
 \lambda_i^+
 &=\int_{-1}^{1}
 \frac{(t+1)(t-1)(t-a)(t-b)(t-s)B_k(t)}
 {(t-\beta_i)(\beta_i^2-1)(\beta_i-a)(\beta_i-b)D^+(\beta_i)} \,d\mu_n(t), \ \ i=1,\ldots,k+1, 
 \label{eq:evenLambdaLagrange}
\end{align}
where $D^+(t)$ is the derivative of $(t-s)B_k(t)$.

\subsection{Positivity of the free-node weights}

Since $(t-s)A_k(t)$ is orthogonal to $\Pi_{k-1}$ with respect to the measure $\nu_{a,b}$, its roots give a Gaussian-type quadrature whose weights are closely related to the free-node weights 
$\lambda_i$, $i=1,\ldots,k+1$, from~\eqref{eq:quadrature}.

\begin{proposition}\label{prop:nuquadrature}
There exist weights $\omega_i>0$, $i=1,\dots,k+1$, such that
\begin{equation}\label{eq:nuquadrature}
 \int_{-1}^{1}u(t)\,d\nu_{a,b}(t) =\sum_{i=1}^{k+1}\omega_i u(\alpha_i)
\end{equation}
for every polynomial $u\in\Pi_{2k}$. 
Moreover,
\begin{equation}\label{eq:lambdaomega}
 \lambda_i=\frac{\omega_i} {(1-\alpha_i)(\alpha_i-a)(\alpha_i-b)}.
\end{equation}
\end{proposition}

\begin{proof}
Interpolation at the roots of $(t-s)A_{k}(t)$ shows exactness through degree $2k$:
the remainder in the integral form~\eqref{eq:remainder} is $(t-s)A_{k}(t)q(t)$ with some $q\in\Pi_{k-1}$. If $\ell_i$ is
the fundamental Lagrange polynomial at $\alpha_i$, then the exactness for $\ell_i^2$ and the positive definiteness up to degree $k$ give
\[ \omega_i=\int_{-1}^{1}\ell_i^2(t) \,d\nu_{a,b}(t)>0. \]
Applying~\eqref{eq:quadrature} to $p(t)=(1-t)(t-a)(t-b)u(t)$ 
we see that the terms at $1,a,b$ vanish. Comparing the result with
\eqref{eq:nuquadrature} gives~\eqref{eq:lambdaomega}.
\end{proof}

\begin{corollary}\label{cor:lambdapositive}
If $\alpha_i\in[-1,a)\cup(b,1)$, then $\lambda_i>0$. In particular, if
$ \alpha_i\in[-1,a)\cup(b,s]=X_{a,b,s} \setminus \{a,b\}$, $1\le i\le k+1$, then every free-node weight $\lambda_i$ is positive.
\end{corollary}

In the even case, we have the following two statements. The proofs are similar to the above. 

\begin{proposition}\label{prop:evenNuQuadrature}
There exist weights $\omega_i^+>0$, $i=1,\ldots,k+1$, such that
\[ 
\int_{-1}^{1}u(t)\,d\nu^+_{a,b}(t)
 =\sum_{i=1}^{k+1}\omega_i^+u(\beta_i)
\]
for every $u\in\Pi_{2k}$. Moreover,
\begin{equation}\label{eq:evenLambdaOmega}
 \lambda_i^+
 =\frac{\omega_i^+}
 {(1-\beta_i^2)(\beta_i-a)(\beta_i-b)}.
\end{equation}
\end{proposition}

\begin{corollary}\label{cor:evenMovingWeightsPositive}
If $\beta_i\in(-1,a)\cup(b,1)$, then $\lambda_i^+>0$. In particular, if $\beta_i\in(-1,a)\cup(b,s]=X_{a,b,s}\setminus\{-1,a,b\}$, $i=1,\ldots,k+1$, then all free-node weights $\lambda_i^+$ are positive.
\end{corollary}

\subsection{The fixed-node weights}

To determine the signs of the fixed-node weights $\rho_1,\rho_a,\rho_b$ we express them explicitly as functions of $a,b,s$ and the moments $M_0,M_1,M_2$. 

Let
\[ m_r:=\int_{-1}^{1}t^r\,d\mu_n(t), \ \ M_r:=m_r- \sum_{i=1}^{k+1}\lambda_i\alpha_i^r, \ \ r=0,1,2. \]
Using $m_0=1$, $m_1=0$ (note that $m_i=0$ for all odd $i$), and $m_2=1/n$, we solve the three equations obtained from~\eqref{eq:quadrature} with $p=1,t,t^2$ to find 
\begin{align}
 \rho_1 &= \frac{M_2-(a+b)M_1+abM_0}{(1-a)(1-b)},
 \label{eq:rhoMoments}\\
 \rho_a &= \frac{M_2-(1+b)M_1+bM_0}{(a-1)(a-b)},
 \label{eq:muaMoments}\\
 \rho_b &= \frac{M_2-(1+a)M_1+aM_0}{(b-1)(b-a)}.
 \label{eq:mubMoments}
\end{align}
Thus, the positivity of the quadrature weights in~\eqref{eq:quadrature} reduces to the finitely many conditions
\[
 \alpha_i\in[-1,a)\cup(b,s], \ \ \rho_1>0, \ \ \rho_a>0, \ \ \rho_b>0,
\]
where these variables are defined by~\eqref{eq:rhoMoments}--\eqref{eq:mubMoments}. This is recorded in the next lemma. 

\begin{lemma} \label{rho-a-b}
If $M_2-(a+b)M_1+abM_0>0$, $M_2-(1+b)M_1+bM_0>0$, and $M_2-(1+a)M_1+aM_0<0$, then $\rho_1>0$, $\rho_a>0$, and $\rho_b>0$, respectively. 
\end{lemma}

\begin{proposition}[Positive odd quadrature]\label{prop:oddPositiveQuadrature}
Assume that
\begin{enumerate}[label=\textup{(\roman*)},leftmargin=2.5em]
 \item $\nu_{a,b}$ is positive definite up to degree $k$;
 \item all roots of $(t-s)A_k(t)$ belong to $(-1,a)\cup(b,s]$;
 \item the numerators in~\eqref{eq:rhoMoments}--\eqref{eq:mubMoments} are positive/negative as in Lemma~\ref{rho-a-b}.
\end{enumerate}
Then all weights in~\eqref{eq:quadrature} are positive.
\end{proposition}

In the even case, the constants $m_r$, $r=0,1,2,3$, are the same, and 
 \[ M_r^+:=m_r-\sum_{i=1}^{k+1}\lambda_i^+\beta_i^r. \]
Solving the four equations obtained from~\eqref{eq:evenQuadrature} by setting consecutively $p(t)=1,t,t^2,t^3$ gives
\begin{align}
 \rho_1^+ &=\frac{
 M_3^++(1-a-b)M_2^++(ab-a-b)M_1^++abM_0^+}
 {2(1-a)(1-b)},
 \label{eq:evenRhoOneMoments}\\
 \rho_{-1}^+ &=\frac{
 -M_3^++(1+a+b)M_2^+-(ab+a+b)M_1^++abM_0^+}
 {2(1+a)(1+b)},
 \label{eq:evenRhoMinusMoments}\\
 \rho_a^+ &=\frac{M_3^+-bM_2^+-M_1^++bM_0^+}
 {(1-a^2)(b-a)},
 \label{eq:evenRhoAMoments}\\
 \rho_b^+ &=\frac{-M_3^++aM_2^++M_1^+-aM_0^+}
 {(1-b^2)(b-a)}.
 \label{eq:evenRhoBMoments}
\end{align}

\begin{lemma}
If all numerators in~\eqref{eq:evenRhoOneMoments}--\eqref{eq:evenRhoBMoments} are positive, then $\rho_1^+>0$, $\rho_{-1}^+>0$, $\rho_a^+>0$, and $\rho_b^+>0$. 
\end{lemma}

\begin{proposition}[Positive even quadrature]\label{prop:evenPositiveQuadrature}
Assume that
\begin{enumerate}[label=\textup{(\roman*)},leftmargin=2.5em]
 \item $\nu^+_{a,b}$ is positive definite up to degree $k$;
 \item all roots of $(t-s)B_k(t)$ belong to $(-1,a)\cup(b,s]$;
 \item the four numerators in
~\eqref{eq:evenRhoOneMoments}--\eqref{eq:evenRhoBMoments} are positive.
\end{enumerate}
Then all weights in~\eqref{eq:evenQuadrature} are positive.
\end{proposition}

\subsection{The quadrature value of the candidate bound}

Since the polynomial $f$ in~\eqref{eq:Fdef} vanishes at
all nodes $a,b,\alpha_1,\ldots,\alpha_{k+1}$ of~\eqref{eq:quadrature} and $\deg f=2k+3$, the quadrature formula for it gives
$ f_0=\rho_1 f(1)$. When $\rho_1>0$, we define the candidate two-point odd Levenshtein bound by
\[
 L_{2k+3}^{a,b}(n,s):=\frac{f(1)}{f_0}=\frac1{\rho_1}.
\]
This is a valid LP bound once the Gegenbauer coefficients
of $f$ are nonnegative.

Similarly, the even (candidate) bound is
\[
 L_{2k+4}^{a,b}(n,s):=\frac1{\rho_1^+}
\]
subject to $\rho_1^+>0$ and positive definiteness of $f_{2k+4}$.

\section{A two-point analogue of the Krein condition and the linear programming bound for maximal \texorpdfstring{$(a,b)$}{(a,b)}-avoiding codes}\label{KreinSec}

Turning to the investigation of the positive definiteness of the candidate polynomials $f$, we start with the $R_i^{a,b}$-expansion of $A_k$. 

\begin{lemma} \label{ak-Rab-pd}
Assume that $a,b,s$ are such that $R_i^{a,b}(s)>0$, $i=0,\dots,k$, and $c_s=-R_{k+1}^{a,b}(s)/R_k^{a,b}(s)>-1$. Then $\gamma_s>0$, $T_k^{a,b}(1,s)>0$, and 
\begin{equation}\label{eq:ApositiveRexpansion}
 A_k(t)=\sum_{i=0}^{k}\theta_iR_i^{a,b}(t) \end{equation}
with $\theta_i>0$, $i=0,\ldots,k$. 
\end{lemma}

\begin{remark} \label{rem6-2}
Let $k\geq 1$.
The conditions $R_i^{a,b}(s)>0$ for every $i=0,\dots,k$ and $c_s=-R_{k+1}^{a,b}(s)/R_k^{a,b}(s)>0$ are satisfied when, for example, $s$ belongs to the interval $(t_k^{a,b},t_{k+1}^{a,b}) \subset (-1,1)$, where $t_i^{a,b}<1$ is the largest root of $R_i^{a,b}$. Note also that $T_k^{a,b}(1,s)>0$ plays a role in Proposition~\ref{prop:kernel}, where we have $d_k>0$ as well. 
\end{remark}

\begin{proof}
We have $\gamma_s=(1-s)/(1+c_s)>0$, 
\[ T_k^{a,b}(1,s)=\sum_{i=0}^{k}r_i^{a,b} R_i^{a,b}(s)>0, \]
and $\theta_i=R_i^{a,b}(s)r_i^{a,b}/T_k^{a,b}(1,s)>0$ (see~\eqref{eq:CDkernel} and~\eqref{eq:AkKernel}). 
\end{proof}

For $c\in\R$ put $H_c(t):=R_{k+1}^{a,b}(t)+cR_k^{a,b}(t)$ and consider the set 
\begin{equation}\label{eq:Cset}
 \mathcal{C}_{n,k}^{a,b}:=\left\{ c\in [0,+\infty):
(t-a)(t-b)R_i^{a,b}(t)H_c(t)\in\Fpos \text{ for every }0\le i\le k \right\}.
\end{equation}
It may happen that the set $\mathcal{C}_{n,k}^{a,b}$ is empty. We are interested in it when it is nonempty and contains $c_s$. This is the analog of what Levenshtein called the Krein condition. For that ordinary Krein condition and, more general, for nonnegative linearization coefficients of Jacobi polynomials, including the Gegenbauer case, see~\cite{Gasper1970II}.

\begin{definition}[$(a,b)$-Krein condition] \label{def:columnKrein}
For fixed $n,k,a,b,s$, we say that the $(a,b)$-Krein condition holds if
\[ c_s=-\frac{R_{k+1}^{a,b}(s)}{R_k^{a,b}(s)} \in \mathcal{C}_{n,k}^{a,b}, \]
in particular 
\[ (t-a)(t-b)R_i^{a,b}(t)\left(R_{k+1}^{a,b}(t)+c_sR_k^{a,b}(t)\right) \in\Fpos \text{ for every } 0\le i\le k. \]
\end{definition}

\begin{proposition}\label{prop:Cinterval}
The set $\mathcal{C}_{n,k}^{a,b}$ is a closed interval, possibly
empty or unbounded.
\end{proposition}

\begin{proof}
Every Gegenbauer coefficient of
\[ (t-a)(t-b)R_i^{a,b}(t)H_c(t)=(t-a)(t-b)R_i^{a,b}(t)R_{k+1}^{a,b}(t)
 +c(t-a)(t-b)R_i^{a,b}(t)R_k^{a,b}(t) \]
is an affine function of $c$. Thus~\eqref{eq:Cset} is the intersection of finitely many closed half-lines.
\end{proof}

A more explicit representation of $\mathcal{C}_{n,k}^{a,b}$ follows. If the coefficient of $P_j^{(n)}$ in the Gegenbauer expansion of 
$(t-a)(t-b)R_i^{a,b}(t)H_c(t)$ is $u_{ij}+c\,v_{ij}$, then
\[ \mathcal{C}_{n,k}^{a,b} =[c_-(a,b),c_+(a,b)], \]
where
\[
 c_-(a,b)
 =\max\left\{0,
 \max_{v_{ij}>0}\left(-\frac{u_{ij}}{v_{ij}}\right)\right\}, \ \ c_+(a,b)
 =\min_{v_{ij}<0}\left(-\frac{u_{ij}}{v_{ij}}\right),
\]
(we additionally restrict to $c\ge0$ in $c_-$; if $v_{ij}=0$, then $u_{ij} \geq 0$ is required, otherwise the set $\mathcal{C}_{n,k}^{a,b}$ is empty) with the usual conventions for empty maxima and minima.

We are now in a position to prove the Levenshtein-type bound, 

\begin{theorem}[Conditional analogue of Levenshtein's Theorem~5.42]
\label{thm:positiveF}
Assume that
\begin{enumerate}[label=\textup{(\roman*)},leftmargin=2.5em]
 \item $\nu_{a,b}$ is positive definite up to degree $k$;
 \item $R_i^{a,b}(s)>0$ for every $i=0,\dots,k$ and $c_s=-R_{k+1}^{a,b}(s)/R_k^{a,b}(s)>-1$; 
 \item the $(a,b)$-Krein condition is satisfied; i.e., $\mathcal{C}_{n,k}^{a,b}$ is non-empty and $c_s\in \mathcal{C}_{n,k}^{a,b}$. 
 \item all weights in~\eqref{eq:quadrature} are nonnegative and $\rho_1>0$.
\end{enumerate}
Then the stationary polynomial $f(t)=(t-a)(t-b)(t-s)A_k^2(t)$
has nonnegative Gegenbauer coefficients, i.e., $f\in\Fpos$, and 
$f_0>0$. Consequently, $f \in \mathcal{LP}_{n,s}^{a,b}$, and the Levenshtein-type bound 
\begin{equation}\label{eq:cardinalityBound}
 \mathcal A(n,s;a,b) \le L_{2k+3}^{a,b}(n,s)=\frac{1}{\rho_1}.
\end{equation}
holds. If equality is attained by an $(a,b)$-avoiding $s$-code $C$, then $I(C)\subset \{a,b,\alpha_1,\ldots,\alpha_{k+1}=s\}$
and $f_iM_i(C)=0$, $i\ge1$.
In particular, if $f_i>0$ for $1\le i\le2k+3$, then $C$ is a
spherical $(2k+3)$-design. If $C$ is an attaining spherical $(2k+3)$-design, it is regular and its distance distribution can be computed from a Vandermonde system.
\end{theorem}

\begin{proof}
By~\eqref{eq:quasiorthogonal} and~\eqref{eq:ApositiveRexpansion}, we write
\[ f(t)=\left[(t-a)(t-b)A_k(t)\right] \cdot \left[(t-s)A_k(t)\right] =\gamma_s\sum_{i=0}^{k}\theta_i (t-a)(t-b)R_i^{a,b}(t)
 \bigl(R_{k+1}^{a,b}(t)+c_sR_k^{a,b}(t)\bigr). \]
Every summand belongs to $\Fpos$, while $\gamma_s>0$ and $\theta_i>0$ by Lemma~\ref{ak-Rab-pd}.
Therefore $f\in\Fpos$. Positive quadrature gives
$f_0=\rho_1 f(1)>0$, whence $f \in \mathcal{LP}_{n,s}^{a,b}$ and the linear programming theorem~\ref{LP-theorem} yields
\eqref{eq:cardinalityBound}.

The equality conditions follow directly from the conditions in Theorem~\ref{LP-theorem}. The calculation of the distance distribution of the attaining $(2k+3)$-design follows from~\eqref{Vand}.
\end{proof}

For $k \geq 1$, under the root location of Remark~\ref{rem6-2}, $s \in (t_k^{a,b},t_{k+1}^{a,b})$ implies the condition (ii) and, in fact, $c_s>0$.

\begin{corollary}
\label{loc-opt}
Assume the hypotheses of Theorem~\ref{thm:positiveF}. Suppose, in addition, that all weights in the quadrature formula~\eqref{eq:quadrature} are positive
and that $\alpha_i\in(-1,a)\cup(b,s), \ \ i=1,\ldots,k$.
Then every polynomial $g(t) \in \mathcal{LP}_{n,s}^{a,b}$ of degree $\deg g \leq 2k+3$ satisfies
\begin{equation}\label{eq:degreeRestrictedOptimality}
 \frac{g(1)}{g_0} \ge \frac1{\rho_1} = L_{2k+3}^{a,b}(n,s).
\end{equation}
In particular, the stationary polynomial $f(t)=(t-a)(t-b)(t-s)A_k^2(t)$ is optimal, up to degree $2k+3$, among all polynomials from $\mathcal{LP}_{n,s}^{a,b}$. Moreover, equality in~\eqref{eq:degreeRestrictedOptimality} holds if and only if
$g(t)=c\,f(t)$
for some constant $c>0$. Thus, $f$ is the unique extremal polynomial
of degree at most $2k+3$, up to multiplication by a positive constant.
\end{corollary}

\begin{proof}
Let $g \in \mathcal{LP}_{n,s}^{a,b} $ have degree at most $2k+3$. Then the quadrature formula~\eqref{eq:quadrature}
applies to $g$ and gives $g_0\le\rho_1g(1)$.
Because $g_0>0$ and $\rho_1>0$,~\eqref{eq:degreeRestrictedOptimality} follows. Moreover, if 
equality holds, then $g$ has the same roots as $f$ counting multiplicity. 
\end{proof}


In Levenshtein's derivation, the even case formulation requires a more involved Krein-type condition. Our exposition allows unification and the even $(a,b)$-Krein condition looks naturally similar to the odd one.

\begin{lemma}\label{lem:evenPositiveRExpansion}
Assume that $a,b,s$ are such that $R_i^{a,b,+}(s)>0$, $i=0,\ldots,k$, and $c_s^+=-R_{k+1}^{a,b,+}(s)/R_k^{a,b,+}(s)>-1$. Then $\gamma_s^+>0$, $T_k^{a,b,+}(1,s)>0$, and
\begin{equation}\label{eq:evenPositiveRExpansion}
 B_k(t)=\sum_{i=0}^{k}\theta_i^+R_i^{a,b,+}(t)
\end{equation}
with $\theta_i^+>0$ for every $i=0,\ldots,k$.
\end{lemma}

\begin{remark}\label{rem:evenRootInterval}
Let $k\geq 1$ and $t_i^{a,b,+}$ denote the largest zero of $R_i^{a,b,+}$. If the relevant
finite-Jacobi zeros lie below $1$, then the usual interlacing argument shows that $s\in(t_k^{a,b,+},t_{k+1}^{a,b,+})$
implies $R_i^{a,b,+}(s)>0$ for $0\le i\le k$ and $c_s^+>0$.
This verifies the sign assumptions of Lemma~\ref{lem:evenPositiveRExpansion},
but it does not by itself imply the even Krein condition below.
\end{remark}

For $c\in\mathbb R$ put $H_c^+(t):=R_{k+1}^{a,b,+}(t)+cR_k^{a,b,+}(t)$
and define
\[
\begin{split}
 \mathcal C_{n,k}^{a,b,+}:=\bigl\{c\in\mathbb R:\ 
 (1+t)(t-a)(t-b)R_i^{a,b,+}(t)H_c^+(t)\in\Fpos \text{ for every }0\le i\le k\bigr\}.
\end{split}
\]

\begin{definition}[Even $(a,b)$-Krein condition]\label{def:evenColumnKrein}
For fixed $n,k,a,b,s$, we say that the even $(a,b)$-Krein condition holds if
\[
 c_s^+
 =-\frac{R_{k+1}^{a,b,+}(s)}{R_k^{a,b,+}(s)}
 \in\mathcal C_{n,k}^{a,b,+}.
\]
\end{definition}

\begin{proposition}\label{prop:evenCinterval}
The set $\mathcal C_{n,k}^{a,b,+}$ is a closed interval, possibly empty or
unbounded.
\end{proposition}

\begin{theorem}[Conditional even analogue of Levenshtein's
Theorem~5.42]\label{thm:evenPositiveF}
Assume that
\begin{enumerate}[label=\textup{(\roman*)},leftmargin=2.5em]
 \item $\nu^+_{a,b}$ is positive definite up to degree $k$;
 \item $R_i^{a,b,+}(s)>0$ for $i=0,\ldots,k$ and $c_s^+>-1$;
 \item the even $(a,b)$-Krein condition holds;
 \item all weights in~\eqref{eq:evenQuadrature} are nonnegative and
 $\rho_1^+>0$.
\end{enumerate}
Then $f_{2k+4}(t)=(1+t)(t-a)(t-b)(t-s)B_k^2(t)$ has nonnegative Gegenbauer coefficients and
$(f_{2k+4})_0>0$; i.e., $f_{2k+4} \in \mathcal{LP}_{n,s}^{a,b}$. Consequently,
\begin{equation}\label{eq:evenCardinalityBound}
 \mathcal A(n,s;a,b)
 \le L_{2k+4}^{a,b}(n,s)
 =\frac1{\rho_1^+}.
\end{equation}
If equality is attained by an $(a,b)$-avoiding $s$-code $C$, then $I(C)\subset \{-1,a,b,\beta_1,\ldots,\beta_{k+1}=s\}$
and $(f_{2k+4})_iM_i(C)=0$, $i\ge1$.
In particular, if $(f_{2k+4})_i>0$ for $1\le i\le2k+4$, then $C$ is a
spherical $(2k+4)$-design. If $C$ is an attaining spherical $(2k+4)$-design, it is regular and its distance distribution can be computed from the Vandermonde system~\eqref{Vand}.
\end{theorem}

\begin{proof}
By~\eqref{eq:evenQuasiorthogonal} and
\eqref{eq:evenPositiveRExpansion},
\begin{align*}
 f_{2k+4}(t)
 &=(1+t)(t-a)(t-b)B_k(t)\bigl((t-s)B_k(t)\bigr)\\
 &=\gamma_s^+
 \sum_{i=0}^{k}\theta_i^+
 (1+t)(t-a)(t-b)R_i^{a,b,+}(t)
 \bigl(R_{k+1}^{a,b,+}(t)+c_s^+R_k^{a,b,+}(t)\bigr).
\end{align*}
Every summand belongs to $\Fpos$, while
$\gamma_s^+>0$ and $\theta_i^+>0$. Hence
$f_{2k+4}\in\Fpos$. The quadrature identity 
\eqref{eq:evenQuadrature} gives
$(f_{2k+4})_0=\rho_1^+f_{2k+4}(1)>0$. The LP theorem yields
\eqref{eq:evenCardinalityBound}, and its equality conditions give the final
statements.
\end{proof}

\begin{corollary}[Degree-restricted optimality in the even branch]
\label{cor:evenDegreeRestrictedOptimality}
Assume the hypotheses of Theorem~\ref{thm:evenPositiveF}, and suppose that all
nontrivial quadrature nodes belong to $X_{a,b,s}$. Then every polynomial
$g\in\mathcal{LP}_{n,s}^{a,b}$ of degree at most $2k+4$ satisfies
\begin{equation}\label{eq:evenDegreeRestrictedOptimality}
 \frac{g(1)}{g_0}\ge\frac1{\rho_1^+}=L_{2k+4}^{a,b}(n,s).
\end{equation}
If, in addition, every quadrature weight is strictly positive and
$\beta_i\in(-1,a)\cup(b,s)$, $\qquad i=1,\ldots,k$, 
then equality in~\eqref{eq:evenDegreeRestrictedOptimality} holds if and only if $g(t)=c\,f_{2k+4}(t)$ for some $c>0$.
\end{corollary}

\begin{remark}\label{rem:evenZeroEndpointWeight}
Strict positivity of $\rho_{-1}^+$ is essential for the uniqueness in Corollary~\ref{cor:evenDegreeRestrictedOptimality}. If
$\rho_{-1}^+=0$, equality in the quadrature argument does not force a zero at $-1$, and an extremal polynomial of degree $2k+3$ may coexist with the even polynomial. The third example in Section~\ref{sec:examples} exhibits exactly this degeneration.
\end{remark}


\section{Universal lower bounds for energy by Hermite interpolation}

\subsection{Linear programming for energy and Hermite interpolation}

Linear programming for energy of codes was introduced in the 1990s by Yudin~\cite{Y1992} (see also~\cite{A1996,A1997,KY1994,KY1997}). Cohn--Kumar~\cite{CK2007} introduced and investigated universally optimal codes, proving that all sharp codes ($m$-distance spherical $(2m-1)$-designs) are universally optimal (see also~\cite{CKM2016,CW2012}). In~\cite{BDHSS2016Energy}, a general linear programming bound (called ULB --- universal lower bound)
was obtained and shown to be attained by all sharp codes (see also~\cite{BDHSS2019Interval,BDHSS2021Lifted}). 

\begin{definition} \label{def-pot}
Given a function $h:[-1,1)\to (-\infty,+\infty)$, continuous on $(-1,1)$, we define the $h$-energy of a spherical code $C$ by 
\[
E_h (C):=\sum_{x\not= y\in C} h(x\cdot y).
\]
\end{definition}

We derive and investigate lower bounds on the quantity 
\[ \mathcal{E}_h(n,N;a,b):=\inf \{E_h(C) : C \subset \mathbb{S}^{n-1} \text{ is $(a,b)$-avoiding}, \ |C|=N\}. \]
We have the following linear programming bounds for $(a,b)$-avoiding codes (not necessarily $s$-codes) of fixed cardinality. 

\begin{theorem}[LP for energy] \label{LP-energy}
Let $n, N \geq 2$ be fixed and $h$ be as in Definition~\ref{def-pot}. Suppose that $f \in \mathbb{R}[t]$ is such that 

{\rm (B1)} $f(t)\le h(t)$ for every $t\in [-1,1) \setminus (a,b)$,

{\rm (B2)} $f_i\ge0$ for every $i\ge1$.

\noindent Then
\[
 \mathcal{E}_h(n,N;a,b)\ge N^2f_0-Nf(1).
\]
If $C \subset \mathbb{S}^{n-1}$ is an $(a,b)$-avoiding code of cardinality $|C|=N$, $f$ is a polynomial satisfying (B1) and (B2) and 
\[ E_h(C)= N^2f_0-Nf(1)=\mathcal{E}_h(n,N;a,b), \]
then $f_iM_i(C)=0$ for every $i \geq 1$ and $f(\langle x,y \rangle)=h(\langle x,y \rangle)$ for every $x,y \in C$, $x \neq y$. 
\end{theorem}

We denote by $\mathcal{LPE}_{n,h}^{a,b}$ the {\it convex admissible set} of polynomials satisfying the conditions (B1) and (B2); i.e., 
\[
 \mathcal{LPE}_{n,h}^{a,b}:=\left\{ f(t)=\sum_{r=0}^{\deg f}f_rP_r^{(n)}(t): f(t)\le h(t)\ \text{for }t\in [-1,1) \setminus (a,b),\ f_r\ge0\ \text{for }r\ge1 \right\}.
\]
Following~\cite{BDHSS2016Energy}, we shall seek optimal, in a sense, polynomials $f \in \mathcal{LPE}_{n,h}^{a,b}$ of degree at most $2k+3$ which are Hermite interpolants to the potential function $h$ at the same Levenshtein nodes from Theorem~\ref{thm:positiveF}. This will provide (B1) while for (B2) we will need conditions similar to the $(a,b)$-Krein condition.

We again explain in detail the odd degree case only. 
Recall that $\alpha_1 < \cdots < \alpha_k$ are the roots of $A_k$ and $\alpha_{k+1}=s$ (the free quadrature nodes); i.e.,
\[ A_k(t)=\kappa (t-\alpha_1) \cdots (t-\alpha_k), \]
where $\kappa$ is a positive constant. 

We use the classical notion of absolute monotonicity (see~\cite{Bernstein1929}) with an one-sided endpoint convention which is specified below.

\begin{definition}
A function $h\in C^\infty([-1,1))$ is absolutely monotone on $[-1,1)$ if $h^{(r)}(t)\ge0$ for every integer $r\ge0$ and every
$t\in[-1,1)$. It is strictly absolutely monotone if all these
inequalities are strict. Derivatives at $-1$ are understood
as right derivatives, continuous from the right. No regularity
or finiteness at $1$ is required.
\end{definition}

Let $h$ be absolutely monotone on $[-1,1)$ and consider the interpolation multiset 
\begin{equation}\label{eq:interpolationMultiset}
\mathcal{T} =\{a,b,\alpha_1,\alpha_1,\ldots,\alpha_k,\alpha_k, s,s\}. \end{equation}
It contains $2k+4$ nodes counting multiplicities. Let $H_h$ be the Hermite
interpolant of $h$ at $\mathcal{T}$ of degree at most $2k+3$; i.e., $H_h(t)$ satisfies
\[ H_h(a)=h(a), \ \ H_h(b)=h(b), \ \ H_h(\alpha_i)=h(\alpha_i), \ \ H_h'(\alpha_i)=h'(\alpha_i), \ \ H_h(s)=h(s), \ \ H_h'(s)=h'(s). \]

The Hermite remainder formula in terms of divided differences with
repeated nodes~\cite{deBoor2005} gives
\[
 h(t)-H_h(t)
 =\frac{1}{\kappa^{2}} h[\mathcal T,t]
 (t-a)(t-b)(t-s)^2A_k^2(t),
\]
where $h[t_0,\ldots,t_r]$ denotes the divided difference of
order $r$, with repeated arguments interpreted in the
Hermite sense; in particular,
$h[t,\ldots,t]=h^{(r)}(t)/r!$ for $r+1$ repeated arguments.
The absolute monotonicity of $h$ implies $h[\mathcal T,t]\ge0$. 
Since $(t-a)(t-b)\ge0$ on $[-1,1) \setminus (a,b)$, we obtain
\begin{equation}\label{eq:minorant}
 H_h(t)\le h(t) \ \mbox{for every} \ t\in [-1,1) \setminus (a,b).
\end{equation}
It remains to prove the positive definiteness of $H_h$.

For the even-degree LP polynomial we write $B_k(t)=\kappa^+(t-\beta_1)\ldots(t-\beta_{k})$, $\kappa^+>0$, and interpolate with the multiset
\begin{equation}\label{eq:evenInterpolationMultiset}
\mathcal T^+:=\{-1,a,b,\beta_1,\beta_1,\ldots,\beta_k,\beta_k, s,s\} \end{equation}
of $2k+5$ elements counted with multiplicity. Now $H_h^+$ is the Hermite interpolant of degree at most $2k+4$ determined by
\[ H_h^+(-1)=h(-1),\ \ H_h^+(a)=h(a),\ \ H_h^+(b)=h(b), \ \ 
H_h^+(s)=h(s),\ \ (H_h^+)'(s)=h'(s), \]
\[ H_h^+(\beta_i)=h(\beta_i),\ \ (H_h^+)'(\beta_i)=h'(\beta_i), \ \ 1\le i\le k. \]
The Hermite remainder is
\[
 h(t)-H_h^+(t) =\frac1{(\kappa^+)^2} h[\mathcal T^+,t]
 (1+t)(t-a)(t-b)(t-s)^2B_k^2(t)
\]
and the absolute monotonicity of $h$ gives $h[\mathcal T^+,t]\ge0$, whence $H_h^+(t)\le h(t)$ for every $t \in [-1,1) \setminus(a,b)$.

\subsection{A direct Newton-positivity hypothesis}

We proceed with the odd case in detail. 

\begin{definition}[Newton positivity]
We say that the interpolation multiset $\mathcal T$ admits a positive Newton
ordering if its elements can be ordered as $\tau_1,\ldots,\tau_{2k+4}$ so that every partial product
\[ PP_r(t):=\prod_{j=1}^{r}(t-\tau_j) \in\Fpos, \ 0\le r\le2k+3. \]
\end{definition}

The next statement is immediate. 

\begin{theorem} \label{Newton-pos}
If $\mathcal T$ admits a positive Newton ordering, then $H_h\in\Fpos$.
\end{theorem}

\begin{proof}
The Newton interpolation formula~\cite{deBoor2005}
\[ H_h(t)=h(\tau_1)+\sum_{r=1}^{2k+3} h[\tau_1,\ldots,\tau_{r+1}]
 PP_r(t) \]
together with the absolute monotonicity of $h$ shows immediately that $H_h\in\Fpos$.
\end{proof}

\subsection{Structural sufficient conditions}

Recall that $\alpha_1<\cdots<\alpha_k<s$ and set
\[ E_r(t):=\prod_{j=1}^{r}(t-\alpha_j), \ \ E_0=1. \]
The following conditions are sufficient for $\mathcal{T}$ to admit a positive Newton ordering.

\begin{enumerate}[label=\textup{(NO\arabic*)},leftmargin=3em]
 \item $a\le0$, so $t-a=P_1^{(n)}(t)-aP_0^{(n)}(t)\in\Fpos$;

 \item every initial root product $E_r$, $r=0,1,\ldots,k$, has a nonnegative $R$-expansion; i.e.,
 \begin{equation}\label{eq:PP}
 E_r(t)=\sum_{j=0}^{r}d_{rj}R_j^{a,b}(t),
 \end{equation}
with $d_{rj}\ge 0$ for every $r=0,\ldots,k$, $j=0,\ldots,r$ (Lemma~\ref{ak-Rab-pd} provides this for $r=k$);

 \item the full $(a,b)$-Krein condition
 \begin{equation}\label{eq:fullSK}
 (t-a)(t-b)R^{a,b}_i(t)R^{a,b}_j(t)\in\Fpos
 \end{equation}
 holds for $0\le i,j\le k$, except possibly for $(i,j)=(k,k)$;
 for $k = 0$ require $(t-a)(t-b) \in \mathcal{F}_{\geq 0}$;

 \item the condition
 \begin{equation}\label{eq:columnSKenergy}
 (t-a)(t-b)R_i^{a,b}(t)
 \bigl(R_{k+1}^{a,b}(t)+c_sR_k^{a,b}(t)\bigr)
 \in\Fpos
 \end{equation}
 holds for $0\le i\le k$.
\end{enumerate}

\begin{lemma}
Assume $k\ge1$. Under \textup{(NO1)}--\textup{(NO4)}, the ordering 
\begin{equation}\label{eq:NewtonOrdering}
 a,b,
 \alpha_1,\alpha_1,
 \ldots,
 \alpha_{k-1},\alpha_{k-1},
 \alpha_k,s,\alpha_k,s.
\end{equation} 
of $\mathcal{T}$ is a positive Newton ordering. 
\end{lemma}

\begin{proof}
The Newton partial products for the proposed ordering are
\[ t-a, \ \ (t-a)(t-b), \]
\[ (t-a)(t-b)E_{r-1}(t)E_r(t), \ \ (t-a)(t-b)E_r^2(t),
 \ \ 1\le r\le k-1, \]
\[ (t-a)(t-b)E_{k-1}(t)E_k(t), \ \ (t-a)(t-b)(t-s)E_{k-1}(t)E_k(t), \]
and finally
\[ (t-a)(t-b)(t-s)E_k^2(t), \]
which is proportional to the Levenshtein-type stationary polynomial $f$ (in fact, $\kappa E_k = A_k$). Conditions~\eqref{eq:PP} and
\eqref{eq:fullSK} control all products not involving $t-s$; the condition~\eqref{eq:columnSKenergy} controls the last two products because
$(t-s)A_k(t)$ is proportional to $R_{k+1}^{a,b}(t)+c_sR_k^{a,b}(t)$.
Thus~\eqref{eq:NewtonOrdering} is a positive Newton ordering.
\end{proof}

\begin{remark}\label{rem:oddNewtonKzero}
For $k=0$, the corresponding ordering is $a,b,s,s$ and the partial products are $t-a$, $(t-a)(t-b)$, and $(t-a)(t-b)(t-s)$. Then the positive Newton ordering conclusion follows from \textup{(NO1)}, the case $i=j=0$ of~\eqref{eq:fullSK}, and the case $i=0$ of~\eqref{eq:columnSKenergy}.
\end{remark}

The Newton positivity is defined analogously in the even case. 
Setting the partial products
\[ 
PP^+_r(t):=\prod_{i=1}^{r}(t-\beta_i), \qquad PP^+_0=1, 
\]
the following conditions are sufficient. 

\begin{enumerate}[label=\textup{(ENO\arabic*)},leftmargin=3.5em]
 \item $a\le0$ (note that this implies $(1+t)(t-a) \in \Fpos$ by the ordinary Krein condition).

 \item Every initial product has a nonnegative expansion in the even
 adjacent basis:
 \[
 PP_r^+(t)=\sum_{j=0}^{r}d_{rj}^+R_j^{a,b,+}(t),
 \qquad d_{rj}^+\ge0,
 \qquad 0\le r\le k.
 \]

 \item The full even $(a,b)$-Krein condition
 \begin{equation}\label{eq:evenFullSK}
 (1+t)(t-a)(t-b)
 R_i^{a,b,+}(t)R_j^{a,b,+}(t)\in\Fpos
 \end{equation}
 holds for $0\le i,j\le k$, except possibly for $(i,j)=(k,k)$.

 \item The even column condition
\begin{equation}\label{eq:evenColumnSKEnergy}
 (1+t)(t-a)(t-b)R_i^{a,b,+}(t)\bigl(R_{k+1}^{a,b,+}(t)+c_s^+R_k^{a,b,+}(t)\bigr) \in\Fpos
 \end{equation}
holds for $0\le i\le k$.
\end{enumerate}

\begin{lemma}\label{lem:evenNewtonOrdering}
Assume $k\ge1$. Under \textup{(ENO1)}--\textup{(ENO4)}, the ordering
\[
 -1,a,b,
 \beta_1,\beta_1,\ldots,
 \beta_{k-1},\beta_{k-1},
 \beta_k,s,\beta_k,s
\]
of $\mathcal T^+$ is a positive Newton ordering.
\end{lemma}

\begin{remark}\label{rem:evenNewtonKzero}
For $k=0$, the corresponding ordering is simply
$-1,a,b,s,s$ and its partial products are $1+t$, $(1+t)(t-a)$, $(1+t)(t-a)(t-b)$, and $(1+t)(t-a)(t-b)(t-s)$. Thus the Newton ordering conclusion follows from \textup{(ENO1)}, the case $i=j=0$ of~\eqref{eq:evenFullSK}, and the case $i=0$ of~\eqref{eq:evenColumnSKEnergy}.
\end{remark}

\subsection{Universal lower bound}

We formulate and prove the odd case universal lower bound on energy, conditioned on $\nu_{a,b}$, proper quadrature nodes placement, and positive Newton ordering.

\begin{theorem}[Energy ULB]
\label{thm:energy}
Let $h$ be absolutely monotone on $[-1,1)$. 
Assume that all weights in the quadrature~\eqref{eq:quadrature} are nonnegative, that all nontrivial nodes belong to $X_{a,b,s}$, and \textup{(NO1)}--\textup{(NO4)} hold. Let $N:=1/\rho_1$ be a positive integer.
Then 
\begin{equation}\label{eq:ULB}
 \mathcal E_h(n,N;a,b) \ge N^2\left(\rho_a h(a)+\rho_b h(b)
 +\sum_{i=1}^{k+1}\lambda_i h(\alpha_i)\right).
\end{equation}
If equality is attained by an $(a,b)$-avoiding code $C$ of cardinality $N$, then the equality conditions in the energy LP theorem hold. In particular, if all Gegenbauer coefficients $(H_h)_i$, $1 \leq i \leq 2k+3$, are strictly positive and the Hermite remainder is strict away from its interpolation nodes, then $C$ is a spherical $(2k+3)$-design and
$I(C)\subset\{a,b,\alpha_1,\ldots,\alpha_{k+1}=s\}$.
If an attaining code $C$ is a spherical $(2k+3)$-design and
$I(C)\subseteq\{a,b,\alpha_1,\ldots,\alpha_{k+1}\}$, then $C$ is distance invariant and, for every $x\in C$, we have
$A_a(x;C)=N\rho_a$, $A_b(x;C)=N\rho_b$, $A_{\alpha_i}(x;C)=N\lambda_i$ for $1\le i\le k+1$.
\end{theorem}

\begin{proof}
We apply Theorem~\ref{LP-energy} with the interpolant $H_h$ as constructed above. By~\eqref{eq:minorant}, we have $H_h\le h$ on the allowed inner-product set. The Newton positivity gives $H_h\in\Fpos$. Then $H_h \in \mathcal{LPE}_{n,h}^{a,b}$ and the energy linear programming Theorem~\ref{LP-energy} yields
\[ 
E_h(C) \ge N^2(H_h)_0-NH_h(1) 
\]
for every $(a,b)$-avoiding code $C \subset \mathbb{S}^{n-1}$ of cardinality $N$ (not necessarily an $s$-code). The quadrature formula~\eqref{eq:quadrature} for $H_h$ and the interpolation conditions give
\[ (H_h)_0 =\rho_1 H_h(1)+\rho_a h(a)+\rho_b h(b)
 +\sum_{i=1}^{k+1}\lambda_i h(\alpha_i). \]
Since $N\rho_1=1$, the $H_h(1)$ terms cancel, giving~\eqref{eq:ULB}.

The equality conditions follow from Theorem~\ref{LP-energy} applying the design properties for the Lagrange basic polynomials associated with the quadrature nodes in the formula~\eqref{eq:quadrature}.
\end{proof}

The even degree bound follows. 

\begin{theorem}[Even degree energy ULB]
\label{thm:evenEnergy}
Let $h$ be absolutely monotone on $[-1,1)$. 
Assume that all weights in the quadrature~\eqref{eq:evenQuadrature} are nonnegative, that all nontrivial nodes belong to $X_{a,b,s}$, and \textup{(ENO1)}--\textup{(ENO4)} hold (with the
interpretation in Remark~\ref{rem:evenNewtonKzero} when $k=0$). Let
$N:=1/\rho_1^+$ be a positive integer. Then
\begin{equation}\label{eq:evenULB}
 \mathcal E_h(n,N;a,b) \ge N^2\Bigg(\rho_{-1}^+h(-1)
+\rho_a^+h(a)+\rho_b^+h(b)+\sum_{i=1}^{k+1}\lambda_i^+h(\beta_i)\Bigg).
\end{equation}
If equality is attained by an $(a,b)$-avoiding code $C$ of cardinality $N$, then the equality conditions in the energy LP theorem hold. In particular, if all Gegenbauer coefficients $(H_h^+)_i$, $1 \leq i \leq 2k+4$, are strictly positive and the Hermite remainder is strict away from its interpolation nodes, then $C$ is a spherical $(2k+4)$-design and
$I(C)\subset\{-1,a,b,\beta_1,\ldots,\beta_{k+1}=s\}$.
If an attaining code $C$ is a spherical $(2k+4)$-design and
$I(C)\subseteq\{-1,a,b,\beta_1,\ldots,\beta_{k+1}\}$, then $C$ is distance invariant and, for every $x\in C$, we have $A_{-1}(x,C)=N\rho_{-1}^+ \in \{0,1\}$,
$A_a(x;C)=N\rho_a^+$, $A_b(x;C)=N\rho_b^+$, $A_{\beta_i}(x;C)=N\lambda_i^+$ for $1\le i\le k+1$.
\end{theorem}

\begin{remark}
The energy bounds remain valid under the weaker assumption
$h^{(r)}\ge0$ for every $r\ge1$, without requiring $h\ge0$.
Indeed, the potentially negative term $h(\tau_1)$ in the
Newton formula influences only the constant Gegenbauer
coefficient, whereas condition (B2) concerns positive degrees.
Thus the interpolant remains energy-admissible, although it
need not belong to $\Fpos$.
Examples include the logarithmic potential
$h(t)=-(1/2)\log(2-2t)$ and the Fejes T\'oth potential
$h(t)=-2(2-2t)^{\alpha/2}$, $0<\alpha<2$, which are not positive on $[-1,1)$ but the conclusions of Theorems~\ref{thm:energy} and~\ref{thm:evenEnergy} hold. 
\end{remark}

\section{Test functions: optimality of the bounds}

Corollary~\ref{loc-opt} gives the optimality of the bound~\eqref{eq:cardinalityBound} for the class of polynomials satisfying the conditions (A1) and (A2) of the linear programming theorem (Theorem~\ref{LP-theorem}) of the same or lower degree. We extend this by proving a necessary and sufficient condition (under certain mild additional assumptions) for global optimality of~\eqref{eq:cardinalityBound}.

Test functions for the Levenshtein bound were defined in 1996~\cite{BDB1996} for the maximal spherical codes problem and (straightforwardly) generalized by Levenshtein for polynomial metric spaces in 1998~\cite{Lev98}. It was shown in 2016 that the same functions work (for an analogous purpose) for universal lower bounds for energy~\cite{BDHSS2016Energy}. 

For every integer $j\ge0$, define test functions
\begin{equation}\label{eq:testfunction}
 \mathcal Q_j^{a,b}(n,s) :=\rho_1+\rho_aP_j^{(n)}(a)
+\rho_bP_j^{(n)}(b)+\sum_{i=1}^{k}\lambda_iP_j^{(n)}(\alpha_i)+\lambda_{k+1}P_{j}^{(n)}(s)
\end{equation}
in the odd case and
\begin{equation}\label{eq:even-testfunction}
 \mathcal Q_j^{a,b,+}(n,s) :=\rho_1^++\rho_{-1}^+(-1)^j+\rho_a^+P_j^{(n)}(a)
+\rho_b^+P_j^{(n)}(b)+\sum_{i=1}^{k}\lambda_i^+P_j^{(n)}(\beta_i)+\lambda_{k+1}^+P_j^{(n)}(s)
\end{equation}
in the even case. Since the quadrature~\eqref{eq:quadrature} is exact through degree $2k+3$, we have
$\mathcal Q_j^{a,b}(n,s) \equiv 0$ for $1\le j\le 2k+3$ (in other words, test functions $\mathcal Q_j^{a,b}(n,s)$ are interesting for $j \geq 2k+4$ only). Similarly for $\mathcal Q_j^{a,b,+}(n,s)$, where the first potentially nonzero is when $j=2k+5$. 

The test functions~\eqref{eq:testfunction}-\eqref{eq:even-testfunction} work both for maximal cardinality codes and minimum energy problems. We consider these two problems separately.

\subsection{Optimality of the maximal code bound \texorpdfstring{\eqref{eq:cardinalityBound}}{!!}}

Recall that $\mathcal{LP}_{n,s}^{a,b}$ is exactly the set of polynomials suitable for an application of Theorem~\ref{LP-theorem}. We consider in detail only the odd degree case. The even-degree case is analogous. 

\begin{theorem} \label{thm:testfunctions}
Assume that all nontrivial quadrature nodes belong to $X_{a,b,s}$ and all quadrature weights are positive.
\begin{enumerate}[label=\textup{(\alph*)},leftmargin=2.5em]
 \item If $\mathcal{Q}_j^{a,b}(n,s)\ge0$ for all positive integers $j\ge 2k+4$, then every $G\in\mathcal{LP}_{n,s}^{a,b}$ satisfies
 \[ \frac{G(1)}{G_0}\ge \frac{1}{\rho_1}. \]
 Thus, the bound~\eqref{eq:cardinalityBound} cannot be improved by linear programming via Theorem~\ref{LP-theorem}.

 \item Assume that the stationary polynomial $f$ has strictly positive Gegenbauer coefficients $f_i$ for all $i=0,1,\ldots,2k+3$. If $\mathcal{Q}_j^{a,b}(n,s)<0$ for some positive integer $j\ge 2k+4$, then there exists
 $G\in\mathcal{LP}_{n,s}^{a,b}$ of degree $j$ such that
 $G(1)/G_0 < 1/\rho_1$, 
 i.e., $G$ gives a better bound on $\mathcal{A}(n,s;a,b)$ than $f$ does. 
\end{enumerate}
\end{theorem}

\begin{proof}
Let
\[
 G(t)=\sum_{r=0}^d G_rP_r^{(n)}(t)
 \in\mathcal{LP}_{n,s}^{a,b}.
\]
Applying the quadrature functional in~\eqref{eq:quadrature} to the Gegenbauer
expansion of $G$ we obtain
\begin{equation}\label{eq:quadratureResidualGeneralG}
\rho_1G(1)+\rho_aG(a)+\rho_bG(b) +\sum_{i=1}^{k+1}\lambda_iG(\alpha_i)
=G_0+\sum_{r=2k+4}^{d}G_r\mathcal Q_r^{a,b}(n,s).
\end{equation}
Indeed, the quadrature functional takes the value $1$ on $P_0^{(n)}$, the
value $0$ on $P_r^{(n)}$ for $1\le r\le 2k+3$, and the value
$\mathcal Q_r^{a,b}(n,s)$ on $P_r^{(n)}$ for $r\ge 2k+4$. 
The left-hand side of~\eqref{eq:quadratureResidualGeneralG} is at most $\rho_1 G(1)$ because every nontrivial node lies
in $X_{a,b,s}$ and $G\le0$ there. Under the assumption in part~(a), the
right-hand side is at least $G_0$. Hence $G_0\le\rho_1 G(1)$,
which completes the proof of (a). 

For part~(b), first divide $P_j^{(n)}$ by $f$, 
$P_j^{(n)}(t)=D_j(t)f(t)+U_j(t)$, where $\deg U_j<2k+3$. 
For a sufficiently large positive constant $c$, we consider the polynomial 
\[ G_c(t):=(D_j(t)+c)f(t)=P_j^{(n)}(t)-U_j(t)+cf(t). \]
Since $X_{a,b,s}$ is compact, $c$ can be chosen so that
$D_j(t)+c\ge0$ on $X_{a,b,s}$, whence $G_c(t)\le0$ there and $(A1)$ will be satisfied by $G_c$ for such $c$. Furthermore, the strict positivity of the coefficients of $f$ permits $c$ to be chosen still larger so that all
Gegenbauer coefficients of $G_c$ are nonnegative and $(G_c)_0>0$.

At every nontrivial quadrature node, we have $U_j=P_j^{(n)}$, since $f$ vanishes there. Exactness of~\eqref{eq:quadrature} for $U_j$ then gives
\[ \mathcal{Q}_j^{a,b}(n,s) =\rho_1+ (U_j)_0-\rho_1 U_j(1). \]
Using $f_0=\rho_1 f(1)$, we obtain
\[ G_c(1)-\frac{(G_c)_0}{\rho_1}=\frac{\mathcal Q_j^{a,b}(n,s)}{\rho_1}<0. \]
Thus $G_c(1)/(G_c)_0<1/\rho_1$, which completes the proof.
\end{proof}

From the proof of Theorem \ref{thm:testfunctions} we conclude the following corollary.

\begin{corollary}\label{cor:testfunctions-codes}
Assume the hypotheses of Theorem~\ref{thm:positiveF}, that all
nontrivial quadrature nodes belong to $X_{a,b,s}$, that all quadrature
weights are positive, and that $f_i>0$ for $0\le i\le2k+3$.
Then $f$ is optimal in the full polynomial Delsarte linear program of Theorem~\ref{LP-theorem} if
and only if
\[ 
\mathcal Q_j^{a,b}(n,s) \ge 0 \ \text{for every }j \ge 2k+4. 
\]
\end{corollary}

We omit the even case results. 

\subsection{Optimality of the ULB \texorpdfstring{\eqref{eq:ULB}}{!!}}

We now formulate the two-node analogue of
\cite[Theorem~4.1]{BDHSS2016Energy}. We set $N=1/\rho_1$, which incorporates the dependence on all parameters. 
We also assume that all weights in the quadrature formula~\eqref{eq:quadrature} are nonnegative, that all its nontrivial nodes belong 
to $X_{a,b,s}$, and that the interpolation multiset $\mathcal T$ from~\eqref{eq:interpolationMultiset} admits a positive Newton ordering. Thus, for every absolutely monotone function $h$, its Hermite interpolant $H_h$ at
$\mathcal T$ satisfies both conditions (B1) and (B2) of Theorem~\ref{LP-energy}; i.e., 
\begin{equation}\label{eq:HuAdmissible}
 H_h(t)\le h(t) \ \text{for every } t \in [-1,1) \setminus (a,b), \ \ H_h\in\Fpos.
\end{equation}
Every polynomial $G\in\mathcal{LPE}_{n,h}^{a,b}$ gives
$\mathcal E_h(n,N;a,b) \ge N^2G_0-NG(1)$. 

Denote the right-hand side of the universal lower bound~\eqref{eq:ULB} by
\[
 \mathcal{R}^{a,b}(n,N;h) :=N^2\left( \rho_a h(a)+\rho_b h(b)
 +\sum_{i=1}^{k+1}\lambda_i h(\alpha_i) \right).
\]
Recall also that the test functions $\mathcal Q_j^{a,b}(n,s)$ are defined in
\eqref{eq:testfunction}, and that $\mathcal Q_j^{a,b}(n,s) \equiv 0$ for every $1\le j\le 2k+3$.

\begin{theorem}[Test functions for the ULB]
\label{thm:testfunctionsULB}
Let $h$ be absolutely monotone on $[-1,1)$.
\begin{enumerate}[label=\textup{(\alph*)},leftmargin=2.5em]
 \item If $\mathcal{Q}_j^{a,b}(n,s)\ge0$ for all positive integers $j\ge 2k+4$, then every $G\in\mathcal{LPE}_{n,h}^{a,b}$ satisfies
 \[ N^2G_0-NG(1) \leq \mathcal{R}^{a,b}(n,N;h);\]
 i.e., the universal lower bound~\eqref{eq:ULB} cannot be improved by any polynomial from $\mathcal{LPE}_{n,h}^{a,b}$.

 \item Assume, in addition, that $h$ is strictly absolutely monotone on $[-1,1)$. If 
 \begin{equation}\label{eq:negativeTestSufficient}
 \mathcal Q_j^{a,b}(n,s)<0
 \end{equation}
 for some $j\ge 2k+4$, then the bound~\eqref{eq:ULB} can be improved by a polynomial from $\mathcal{LPE}_{n,h}^{a,b}$ of degree exactly $j$. More precisely, for every sufficiently small $\varepsilon>0$, put
 \[
 :=h(t)-\varepsilon P_j^{(n)}(t)
 \]
 and let $H_{\widetilde h_{\varepsilon,j}}$ be its Hermite interpolant at $\mathcal T$. Then
 \begin{equation}\label{eq:improvingEnergyPolynomial}
 F_{\varepsilon,j}(t) :=\varepsilon P_j^{(n)}(t)
 +H_{\widetilde h_{\varepsilon,j}}(t)
 \end{equation}
 belongs to $\mathcal{LPE}_{n,h}^{a,b}$, has degree $j$, and gives the bound
 \begin{equation}\label{eq:quantitativeULBImprovement}
 \mathcal E_h(n,N;a,b)
 \ge
 \mathcal{R}^{a,b}(n,N;h)
 -\varepsilon N^2\mathcal Q_j^{a,b}(n,s)
 >\mathcal{R}^{a,b}(n,N;h).
 \end{equation}
\end{enumerate}
Thus, for strictly absolutely monotone potentials, the ULB can be improved by
an LP polynomial of degree at least $2k+4$ if and only if
$\mathcal Q_j^{a,b}(n,s)<0$ for some $j\ge 2k+4$.
\end{theorem}

\begin{proof}
Let
\[ G(t)=\sum_{r=0}^{d}G_rP_r^{(n)}(t)
 \in\mathcal{LPE}_{n,h}^{a,b}. \]
We have~\eqref{eq:quadratureResidualGeneralG} again and, since $N\rho_1=1$, it can be rewritten as
\[
N^2G_0-NG(1)=N^2\Bigg(\rho_aG(a)+\rho_bG(b)
 +\sum_{i=1}^{k+1}\lambda_iG(\alpha_i)
 -\sum_{r=2k+4}^{d}G_r\mathcal Q_r^{a,b}(n,s)\Bigg).
\]
All quadrature weights are positive, all nontrivial quadrature nodes belong to
$X_{a,b,s}$, and $G\le h$ on $[-1,1) \setminus (a,b)$. Hence,
\[
 N^2G_0-NG(1)
 \le
 \mathcal{R}^{a,b}(n,N;h)
 -N^2\sum_{r=2k+4}^{d}G_r\mathcal Q_r^{a,b}(n,s) \leq \mathcal{R}^{a,b}(n,N;h),
\]
which proves part~(a).

For part~(b), fix $j\ge 2k+4$ satisfying
\eqref{eq:negativeTestSufficient}. Since $h^{(r+1)}\ge0$, each derivative $h^{(r)}$ is
nondecreasing on $[-1,1)$. Hence, for $0\le r\le j$, we have 
$ \delta_r:=h^{(r)}(-1)>0$ and, therefore, $h^{(r)}(t)\ge\delta_r$ for every $t \in [-1,1)$. Set
\[
 M_r:=\max_{t\in[-1,1]}
 \left|\left(P_j^{(n)}\right)^{(r)}(t)\right|,
 \qquad 0\le r\le j
\]
and choose
\[
 0<\varepsilon\le
 \min_{\substack{0\le r\le j, \ M_r>0}}
 \frac{\delta_r}{M_r}.
\]
Then, for $0\le r\le j$ and $t\in[-1,1)$, we have 
$\left(\widetilde h_{\varepsilon,j}\right)^{(r)}(t)
 \ge \delta_r-\varepsilon M_r\ge0$,
while for $r>j$ the derivatives are unchanged. Thus, $\widetilde h_{\varepsilon,j}$ is absolutely monotone on $[-1,1)$.

By~\eqref{eq:HuAdmissible}, its Hermite interpolant to $\widetilde h_{\varepsilon,j}$ satisfies
\[ H_{\widetilde h_{\varepsilon,j}}(t)
 \le\widetilde h_{\varepsilon,j}(t)
 \ \ t\in [-1,1) \setminus (a,b), \ \
 H_{\widetilde h_{\varepsilon,j}}\in\Fpos. \]
It follows from~\eqref{eq:improvingEnergyPolynomial} that
$F_{\varepsilon,j}(t)\le h(t)$ for every $t\in [-1,1) \setminus (a,b)$. Moreover, adding $\varepsilon P_j^{(n)}$ preserves the non-negativity of all
Gegenbauer coefficients of positive degree. Since
$\deg H_{\widetilde h_{\varepsilon,j}}\le 2k+3<j$, the polynomial
$F_{\varepsilon,j}$ has degree exactly $j$. Therefore
$F_{\varepsilon,j}\in\mathcal{LPE}_{n,h}^{a,b}$.

It remains to calculate the resulting lower bound. By the interpolation conditions, we have $H_{\widetilde h_{\varepsilon,j}}(a)
 =\widetilde h_{\varepsilon,j}(a)$, $H_{\widetilde h_{\varepsilon,j}}(b)=\widetilde h_{\varepsilon,j}(b)$,
and $H_{\widetilde h_{\varepsilon,j}}(\alpha_i)
 =\widetilde h_{\varepsilon,j}(\alpha_i)$ for every $1\le i\le k+1$. Since $H_{\widetilde h_{\varepsilon,j}}$ has degree at most $2k+3$, the exact quadrature formula gives
\begin{equation}\label{eq:HtildeQuadrature}
 \left(H_{\widetilde h_{\varepsilon,j}}\right)_0
 =\rho_1H_{\widetilde h_{\varepsilon,j}}(1)
 +\rho_a\widetilde h_{\varepsilon,j}(a)
 +\rho_b\widetilde h_{\varepsilon,j}(b)+\sum_{i=1}^{k+1}
 \lambda_i\widetilde h_{\varepsilon,j}(\alpha_i).
\end{equation}
The polynomial $P_j^{(n)}$ has zero constant Gegenbauer coefficient and
$P_j^{(n)}(1)=1$. Using $N\rho_1=1$,~\eqref{eq:HtildeQuadrature}, and the
definition~\eqref{eq:testfunction}, we obtain
\begin{align*}
 N^2\left(F_{\varepsilon,j}\right)_0-NF_{\varepsilon,j}(1) 
 &= N^2\left(H_{\widetilde h_{\varepsilon,j}}\right)_0
 -N\left(H_{\widetilde h_{\varepsilon,j}} (1)+\varepsilon\right)\\
 &= N^2\left(\rho_a\widetilde h_{\varepsilon,j}(a)
 +\rho_b\widetilde h_{\varepsilon,j}(b)
 +\sum_{i=1}^{k+1}
 \lambda_i\widetilde h_{\varepsilon,j}(\alpha_i)
 \right)-N\varepsilon\\
 &= \mathcal{R}^{a,b}(n,N;h)-\varepsilon N^2\left(
 \rho_1+\rho_aP_j^{(n)}(a)+\rho_bP_j^{(n)}(b)
 +\sum_{i=1}^{k+1}\lambda_iP_j^{(n)}(\alpha_i)
 \right)\\
 &= \mathcal{R}^{a,b}(n,N;h)-\varepsilon N^2\mathcal Q_j^{a,b}(n,s).
\end{align*}
The last expression is strictly larger than
$\mathcal{R}^{a,b}(n,N;h)$ by~\eqref{eq:negativeTestSufficient}. Applying the energy
LP bound proves~\eqref{eq:quantitativeULBImprovement} and completes the proof.
\end{proof}

The proof of Theorem \ref{thm:testfunctionsULB} implies the next corollary.

\begin{corollary}\label{cor:testfunctionsULB}
Under the hypotheses of Theorem~\ref{thm:testfunctionsULB}, suppose that $h$ is
strictly absolutely monotone on $[-1,1)$. Then the Hermite-interpolation ULB
\eqref{eq:ULB} is optimal in the full polynomial Delsarte--Yudin linear program
if and only if
\[ \mathcal Q_j^{a,b}(n,s)\ge0 \ \text{for every }j\ge2k+4. \]
\end{corollary}

Under the combined hypotheses of Corollaries~\ref{cor:testfunctions-codes} and~\ref{cor:testfunctionsULB}, including strict positivity of the Gegenbauer coefficients of the stationary cardinality
polynomial and strict absolute monotonicity of $h$, the cardinality bound
and the Hermite-interpolation energy bound are simultaneously optimal
in their respective full polynomial linear programmes, or both admit
a strict polynomial improvement.

The even case of the test functions theorem and corollary are analogous.

\section{Three examples, odd and even degree} \label{sec:examples}

Throughout this section, $h$ is absolutely monotone on $[-1,1)$ with
 the endpoint convention of Section~7. The energy bounds apply to all
$(a,b)$-avoiding codes of the indicated cardinality, without an upper
cutoff on their inner products. We use the ordered-pair convention for
$E_h$. Distance distributions are listed in increasing order of inner
products and give the number of neighbours of each point. In each example,
these multiplicities are $N$ times the corresponding nontrivial
quadrature weights; consequently, they sum to $N-1$.

\subsection{A 23-dimensional odd example with \texorpdfstring{$k=2$}{k=2}} \label{sec47104}
Take
\[ n=23, \ \ k=2, \ \ a=-\frac35, \ \ b=-\frac13, \ \ s=\frac7{15}. \]
This is one of the exact $(a,b)$-avoiding examples in
\cite[Theorem~5.1]{BCD2026} of cardinality $N=47104$.

The modified measure is
\[ d\nu_{a,b}(t) =(1-t)\left(t+\frac35\right)
 \left(t+\frac13\right)d\mu_{23}(t). \]
The first normalized orthogonal polynomials are (we omit the superscripts $a,b$ in the rest of the example):
\[ R_0(t)=1,\ \ R_1(t)=\frac{175t-23}{152}, \ \ R_2(t)=\frac{1425t^2-430t-7}{988}, \]
\[ R_3(t)= \frac{5481t^3-3471t^2+323t+35}{2368}. \]
Their squared norms begin with
\[ \frac{1}{r_0^{a,b}}=\frac{14}{69}>0, \ \ \frac{1}{r_1^{a,b}}=\frac{77}{10488}>0, \ \ \frac{1}{r_2^{a,b}}=\frac{154}{369265}>0, \]
while
\[ \int_{-1}^1 R_3^2(t)\,d\nu_{a,b}(t) =-\frac{1463}{31487000}<0. \]
Thus positivity holds exactly as far as needed for $k=2$.

We obtain
\[
 c_s=-\frac{R_3(s)}{R_2(s)}=\frac{247}{4625}
\]
and
\[
 A_2(t)
 =\frac{T_2^{a,b}(t,s)}{T_2^{a,b}(1,s)}
 =\frac{(5t-1)(15t+1)}{64}.
\]
Therefore $(t-s)A_2(t)$ has the roots $-1/15$, $1/5$, and $7/15$.
The exact quadrature is
\[
\begin{aligned}
p_0=\frac1{47104}\Bigg[p(1)+275p\left(-\frac35\right)
 +7128p\left(-\frac13\right)+22275p\left(-\frac1{15}\right)
+15400p\left(\frac15\right)+2025p\left(\frac7{15}\right)\Bigg]
\end{aligned}
\]
for every $p\in\Pi_7$. All weights are positive and
$ \rho_1=1/47104$. The candidate polynomial is
\[ f(t) =\left(t+\frac35\right) \left(t+\frac13\right)
\left(t-\frac7{15}\right) \left(\frac{(5t-1)(15t+1)}{64}\right)^2.
\]
Consequently,
\[ \frac{f(1)}{f_0}=47104. \]
The cited $T$-avoiding paper~\cite{BCD2026} verifies the required Gegenbauer positivity and therefore obtains the sharp bound
\[ \mathcal A\left(23,\frac7{15};-\frac35,-\frac13\right)
 \le47104. \]
The $(a,b)$-Krein condition can also be verified explicitly. Direct
Gegenbauer expansion of the three columns gives
\[
 \mathcal C_{23,2}^{-3/5,-1/3}
 =\left[\frac{767}{281200},+\infty\right),
 \qquad c_s=\frac{247}{4625}>\frac{767}{281200}.
\]
Moreover,
\[
 R_0(s)=1,\qquad R_1(s)=\frac{22}{57}>0,
 \qquad R_2(s)=\frac{77}{741}>0.
\]
Thus the positivity of $f$ also follows from the $(a,b)$-Krein criterion.

The bound is attained by the spherical $7$-design
$C_{1/4}\subset\mathbb S^{22}$ described in~\cite{BCD2026}. Its inner-product
set and distance distribution are
\[
 I(C_{1/4})=\left\{-\frac35,-\frac13,-\frac1{15},
 \frac15,\frac7{15}\right\},
 \qquad
 F(C_{1/4})=(275,7128,22275,15400,2025).
\]
For energy, use the Newton ordering
\[
 -\frac35,-\frac13,-\frac1{15},-\frac1{15},
 \frac15,\frac7{15},\frac15,\frac7{15}.
\]
Direct expansion shows that every required partial product, of degrees
$0,\ldots,7$, has positive Gegenbauer coefficients. The degree-$7$
Hermite interpolant is therefore an admissible minorant. The quadrature
and $N\rho_1=1$ give
\[
\begin{aligned}
 E_h(C)\ge{}&
 \mathcal E_h\left(23,47104;-\frac35,-\frac13\right)\\
 ={}&47104\Bigg[275h\left(-\frac35\right)
 +7128h\left(-\frac13\right)
 +22275h\left(-\frac1{15}\right) +15400h\left(\frac15\right)
 +2025h\left(\frac7{15}\right)\Bigg]
\end{aligned}
\]
for every $(-3/5,-1/3)$-avoiding code $C\subset\mathbb S^{22}$ of
cardinality $47104$. Equality holds for $C_{1/4}$, recovering the energy
bound of~\cite[Theorem~7.1]{BCD2026}.

\subsection{A 23-dimensional odd example with \texorpdfstring{$k=1$}{k=1}} \label{sec11178}

Take
\[
 n=23,\qquad k=1,\qquad
 a=-\frac12,\qquad b=-\frac15,\qquad s=\frac25.
\]
This is one of the exact $(a,b)$-avoiding examples in
\cite[Theorem~5.3]{BCD2026} of cardinality $N=11178$.

The modified measure is
\[
 d\nu_{a,b}(t)
 =(1-t)\left(t+\frac12\right)\left(t+\frac15\right)d\mu_{23}(t).
\]
The first normalized orthogonal polynomials are (we omit the superscripts
$a,b$ in the rest of the example):
\[
 R_0(t)=1,\qquad
 R_1(t)=\frac{65t-12}{53},\qquad
 R_2(t)=\frac{1575t^2-820t+69}{824}.
\]
Their squared norms begin with
\[
 \frac1{r_0^{a,b}}=\frac{13}{115}>0,
 \qquad
 \frac1{r_1^{a,b}}=\frac{1001}{323035}>0,
\]
while
\[
 \int_{-1}^1 R_2^2(t)\,d\nu_{a,b}(t)
 =-\frac{1001}{4880140}<0.
\]
Thus positivity holds exactly as far as needed for $k=1$.

At $s=2/5$ we have
\[
 R_1(s)=\frac{14}{53},\qquad
 R_2(s)=-\frac7{824},\qquad
 c_s=-\frac{R_2(s)}{R_1(s)}=\frac{53}{1648}.
\]
We obtain
\[
 A_1(t)
 =\frac{T_1^{a,b}(t,s)}{T_1^{a,b}(1,s)}
 =\frac{10t-1}{9}.
\]
Therefore $(t-s)A_1(t)$ has the roots $1/10$ and $2/5$.
The exact quadrature is
\[
\begin{aligned}
 p_0=\frac1{11178}\Bigg[p(1)
 +352p\left(-\frac12\right)
 +4125p\left(-\frac15\right)+5600p\left(\frac1{10}\right)
 +1100p\left(\frac25\right)\Bigg]
\end{aligned}
\]
for every $p\in\Pi_5$. All weights are positive and
$\rho_1=1/11178$. The candidate polynomial is
\[
 f(t)
 =\left(t+\frac12\right)\left(t+\frac15\right)
 \left(t-\frac25\right)\left(\frac{10t-1}{9}\right)^2.
\]
Consequently,
\[
 f_0=\frac1{10350},\qquad
 f(1)=\frac{27}{25},\qquad
 \frac{f(1)}{f_0}=11178.
\]

In this example, the $(a,b)$-Krein condition can also be verified
explicitly. Expanding the two columns
\[
 \left(t+\frac12\right)\left(t+\frac15\right)
 R_i(t)\bigl(R_2(t)+cR_1(t)\bigr),\qquad i=0,1,
\]
in the Gegenbauer basis gives
\[
 \mathcal C_{23,1}^{-1/2,-1/5}
 =\left[\frac{2279}{278512},+\infty\right),
 \qquad
 c_s=\frac{53}{1648}>\frac{2279}{278512}.
\]
Since $R_0(s)=1$ and $R_1(s)=14/53>0$, the $(a,b)$-Krein criterion
therefore yields the required Gegenbauer positivity of $f$ and the sharp
bound
\[
 \mathcal A\left(23,\frac25;-\frac12,-\frac15\right)\le11178.
\]
Equality is attained by the spherical $5$-design $C_{\sqrt6/6}$ described
in~\cite{BCD2026}.
Its inner-product set and distance distribution are
\[
 I(C_{\sqrt6/6})=\left\{-\frac12,-\frac15,\frac1{10},\frac25\right\},
 \qquad F(C_{\sqrt6/6})=(352,4125,5600,1100).
\]
The Newton ordering
\[
 -\frac12,-\frac15,\frac1{10},\frac25,\frac1{10},\frac25
\]
has positive Gegenbauer coefficients in every required partial product,
of degrees $0,\ldots,5$. Hence the degree-$5$ Hermite interpolant is
energy-admissible, and
\[
\begin{aligned}
 E_h(C)\ge{}&
 \mathcal E_h\left(23,11178;-\frac12,-\frac15\right)\\
 ={}&11178\Bigg[352h\left(-\frac12\right)
 +4125h\left(-\frac15\right) +5600h\left(\frac1{10}\right)
 +1100h\left(\frac25\right)\Bigg]
\end{aligned}
\]
for every $(-1/2,-1/5)$-avoiding code $C\subset\mathbb S^{22}$ of
cardinality $11178$. Equality holds for $C_{\sqrt6/6}$, in agreement
with~\cite[Theorem~7.4]{BCD2026}.

\subsection{A degenerate 22-dimensional even example}\label{sec2025}

There is an exact degree-$4$ certificate in the $T$-avoiding examples of
\cite{BCD2026} of cardinality $N=2025$. Take
\[
 n=22,
 \qquad
 a=-\frac4{11},
 \qquad
 b=-\frac1{44},
 \qquad
 s=\frac7{22},
 \qquad k=0.
\]
The even stationary polynomial is
\begin{equation}\label{eq:even2025Polynomial}
 f_4(t)
 =(1+t)\left(t+\frac4{11}\right)
 \left(t+\frac1{44}\right)
 \left(t-\frac7{22}\right).
\end{equation}
Its Gegenbauer expansion in dimension $22$ is
\begin{equation}\label{eq:even2025Expansion}
 f_4(t) =\frac5{5324}P_0^{(22)}(t) +\frac{691}{42592}P_1^{(22)}(t)
 +\frac{48699}{276848}P_2^{(22)}(t)+\frac{329}{352}P_3^{(22)}(t)
 +\frac{161}{208}P_4^{(22)}(t),
\end{equation}
so all coefficients are positive. The exact quadrature is
\[
\begin{aligned}
 p_0=0\cdot p(-1)+\frac1{2025}p(1)
 +\frac{22}{135}p\left(-\frac4{11}\right)+\frac{1232}{2025}p\left(-\frac1{44}\right)
 +\frac{154}{675}p\left(\frac7{22}\right),
 \qquad p\in\Pi_4.
\end{aligned}
\]
Thus $\rho_{-1}^+=0$ and
\[ \frac{f_4(1)}{(f_4)_0}=2025. \]
However, this is not a genuinely even improvement. The cubic polynomial
\[
 q_3(t) :=\frac{f_4(t)}{1+t}
 =\left(t+\frac4{11}\right) \left(t+\frac1{44}\right)
 \left(t-\frac7{22}\right)
\]
already has the positive Gegenbauer expansion
\[
 q_3(t) =\frac5{10648}P_0^{(22)}(t)
 +\frac5{484}P_1^{(22)}(t) +\frac{63}{968}P_2^{(22)}(t)
 +\frac78P_3^{(22)}(t),
\]
and gives the same quotient $2025$. The zero weight at $-1$ is exactly the quadrature manifestation of the junction with the neighboring odd branch.

For completeness, the even modified measure in this example has
\[
 \int_{-1}^{1}d\nu^+_{a,b}(t)=\frac{1015}{21296}>0,
\]
and its first normalized orthogonal polynomial is
\[
 R_1^{a,b,+}(t)=\frac{580t-187}{393}.
\]
At $s=7/22$ one has
\[
 R_1^{a,b,+}(s)=-\frac9{1441},
 \qquad
 c_s^+=\frac9{1441}.
\]

The cardinality bound
\[
 \mathcal A\left(22,\frac7{22};-\frac4{11},-\frac1{44}\right)\le2025
\]
is attained by the spherical $4$-design $C_{2025}\subset\mathbb S^{21}$
from~\cite{BCD2026}. Its inner-product set and distance distribution are
\[
 I(C_{2025})=\left\{-\frac4{11},-\frac1{44},\frac7{22}\right\},
 \qquad F(C_{2025})=(330,1232,462).
\]
In particular, the zero quadrature weight at $-1$ corresponds to the
absence of antipodal pairs.

For energy, the even Newton ordering is
\[
 -1,-\frac4{11},-\frac1{44},\frac7{22},\frac7{22}.
\]
Its first three nonconstant partial products are products of positive
definite linear factors, and the fourth is $f_4$, whose Gegenbauer
coefficients are positive by~\eqref{eq:even2025Expansion}. Thus the
corresponding degree-$4$ Hermite interpolant is energy-admissible.
The quadrature cancellation remains valid when $\rho_{-1}^+=0$:
only nonnegative weights and $N\rho_1^+=1$ are needed. It gives
\[
\begin{aligned}
 E_h(C)\ge{}&
 \mathcal E_h\left(22,2025;-\frac4{11},-\frac1{44}\right)\\
 ={}&2025\Bigg[330h\left(-\frac4{11}\right)
 +1232h\left(-\frac1{44}\right)
 +462h\left(\frac7{22}\right)\Bigg]
\end{aligned}
\]
for every $(-4/11,-1/44)$-avoiding code $C\subset\mathbb S^{21}$ of
cardinality $2025$. Equality holds for $C_{2025}$.

The same energy bound follows from the cubic Hermite interpolant at
$a,b,s,s$. Indeed, the ordering $a,b,s,s$ has partial products
$t-a$, $(t-a)(t-b)$, and $q_3(t)$, all with positive Gegenbauer
coefficients. Its remainder is
\[
 h(t)-H_h^{(3)}(t)
 =h[a,b,s,s,t](t-a)(t-b)(t-s)^2\ge0,
 \qquad t\in[-1,1)\setminus(a,b).
\]
Thus the degeneration also appears in the energy problem: the cubic
and quartic interpolants yield the same bound. This recovers
\cite[Theorem~7.9]{BCD2026}.

\bigskip

{\bf Acknowledgments.} The authors used OpenAI ChatGPT (GPT-5.6 Sol, Astra) to discuss proof strategies,
check calculations, and to assist with the manuscript preparation. All mathematical
statements, proofs, computations, and citations were independently verified by the
authors, who take full responsibility for the content of the manuscript. PB and DC are supported, in part, by Bulgarian NSF grant KP06-Austria/8-2025. PD is supported, in part, by the Lilly Endowment. PG's research was supported by the Austrian Academic Exchange Agency (OeAD), BG 07/2025. PG's research was funded in whole, or in part, by the Austrian Science Fund (FWF) 10.55776/I6750.

\section*{Appendix }
\label{sec:researchProgramme}

For fixed $n\ge2$, $k\ge0$, and $-1\le a<b<s<1$, we develop a programme for derivation and optimality investigation of the cardinality and energy bounds of Theorems~\ref{thm:positiveF} (\ref{thm:evenPositiveF}) and~\ref{thm:energy} (\ref{thm:evenEnergy}). The odd and even branches have
degrees $m=2k+3$ and $m=2k+4$, respectively; in the even branch we
also require $a>-1$. Write $f_m$ for the corresponding stationary
polynomial in~\eqref{f-poly}, and put
\[
 w_1:=
 \begin{cases}
 \rho_1,&m=2k+3,\\
 \rho_1^+,&m=2k+4.
 \end{cases}
\]
We consider the construction of the bound, its optimality in the polynomial
linear programme, and its attaining by a spherical code as separate questions. Then we describe the energy bounds similarly. 

\subsection*{Construction and optimality of the cardinality bounds}

\begin{enumerate}[leftmargin=2.2em]
\item \textbf{Establish truncated positive definiteness.}
Prove that $\nu_{a,b}$ in the odd branch, or $\nu_{a,b}^+$ in the
even branch, is positive definite up to degree $k$, using the
corresponding Gram matrix. Construct the orthogonal polynomials $R_i^{a,b}$ or $R_i^{a,b,+}$
through degree $k+1$, verify that their values at $1$ are nonzero,
and normalize them to equal $1$ there. 

\item \textbf{Construct the stationary polynomial and locate its roots.}
Using the corresponding Christoffel--Darboux kernel, compute
\[
\begin{aligned}
 c_s&=-\frac{R_{k+1}^{a,b}(s)}{R_k^{a,b}(s)},
 &A_k(t)&=\frac{T_k^{a,b}(t,s)}{T_k^{a,b}(1,s)},\\
 c_s^+&=-\frac{R_{k+1}^{a,b,+}(s)}{R_k^{a,b,+}(s)},
 &B_k(t)&=\frac{T_k^{a,b,+}(t,s)}{T_k^{a,b,+}(1,s)},
\end{aligned}
\]
after verifying that the relevant denominators are nonzero.
Establish that the roots are simple, that $s$ is the largest root,
and that they satisfy
\[
\begin{aligned}
 &\alpha_1<\cdots<\alpha_{k+1}=s,
 &&\alpha_i\in[-1,a)\cup(b,s]
 &&\text{in the odd branch},\\
 &\beta_1<\cdots<\beta_{k+1}=s,
 &&\beta_i\in(-1,a)\cup(b,s]
 &&\text{in the even branch}.
\end{aligned}
\]
These conditions also ensure that the free nodes are distinct
from the fixed nodes. Real-rootedness follows from the finite
Jacobi-matrix representation, but the required root locations
must be verified separately. For $k=0$, use $A_0=B_0=1$ and
$\Pi_{-1}=\{0\}$; no largest-root notation for $R_0$ is needed.

\item \textbf{Determine the signs of the quadrature weights.}
The quadrature formulas~\eqref{eq:quadrature} and~\eqref{eq:evenQuadrature}
are exact through degree $m$. The root locations in the preceding
step imply non-negativity of all free-node weights by
\eqref{eq:lambdaomega} and~\eqref{eq:evenLambdaOmega}.
Verify the remaining conditions
\[
\begin{array}{ll}
 \rho_1>0,\quad \rho_a,\rho_b\ge0,
 &\text{in the odd branch},\\[2pt]
 \rho_1^+>0,\quad \rho_{-1}^+,\rho_a^+,\rho_b^+\ge0,
 &\text{in the even branch},
\end{array}
\]
using~\eqref{eq:rhoMoments}--\eqref{eq:mubMoments} or
\eqref{eq:evenRhoOneMoments}--\eqref{eq:evenRhoBMoments}.
If one is interested in the uniqueness, then record which inequalities are strict. 

\item \textbf{Prove Gegenbauer positivity.}
A structural sufficient criterion is to verify
$R_i^{a,b}(s)>0$ for $0\le i\le k$, $c_s>-1$, and
$c_s\in\mathcal C_{n,k}^{a,b}$ in the odd branch, or
$R_i^{a,b,+}(s)>0$ for $0\le i\le k$, $c_s^+>-1$, and
$c_s^+\in\mathcal C_{n,k}^{a,b,+}$ in the even branch.
Theorems~\ref{thm:positiveF} and~\ref{thm:evenPositiveF} then give
$f_m\in\Fpos$. Alternatively, verify its Gegenbauer coefficients by
direct computation; the Krein conditions are sufficient, not necessary,
for this conclusion. Since
\[
 (f_m)_0=w_1f_m(1)>0,
\]
the obtained polynomial is LP-admissible and yields
\[
 \mathcal A(n,s;a,b)\le\frac{f_m(1)}{(f_m)_0}=\frac1{w_1}.
\]

\item \textbf{Establish optimality up to degree $m$.}
For every $g\in\mathcal{LP}_{n,s}^{a,b}$ with $\deg g\le m$,
exactness and nonnegative quadrature weights give
\[
 g_0\le w_1g(1),\qquad \frac{g(1)}{g_0}\ge\frac1{w_1}.
\]
If all quadrature weights are strictly positive and the roots of
$A_k$ or $B_k$ lie in $(-1,a)\cup(b,s)$, equality forces
$g$ to be a positive multiple of $f_m$, as in
Corollaries~\ref{loc-opt} and~\ref{cor:evenDegreeRestrictedOptimality}.
Zero weights do not invalidate the bound or degree-restricted
optimality, but uniqueness must then be considered separately.
In particular, $\rho_{-1}^+=0$ can allow an odd polynomial to attain
the same bound as the even polynomial; see
Remark~\ref{rem:evenZeroEndpointWeight}.

\item \textbf{Test optimality over all polynomial degrees.}
Let $\mathcal Q_j$ denote~\eqref{eq:testfunction} in the odd branch
and~\eqref{eq:even-testfunction} in the even branch. Exactness of the quadrature formulas then gives
$\mathcal Q_0=1$ and $\mathcal Q_j=0$ for $1\le j\le m$.
If
\[
 \mathcal Q_j\ge0\qquad\text{for every }j\ge m+1,
\]
then the bound is optimal in the full polynomial linear programme.
Conversely, if every Gegenbauer coefficient $(f_m)_i$ is strictly
positive for $0\le i\le m$, a negative value $\mathcal Q_j<0$
produces an improving polynomial of degree $j$.
Thus, under this strict coefficient assumption, the test-function
condition is both necessary and sufficient. 
Note that, for $n\ge3$, let $w_{-1}$ be the total quadrature weight at $-1$, with
$w_{-1}=0$ if that node is absent. Since
$P_j^{(n)}(t)\to0$ for every fixed $t\in(-1,1)$ as $j\to \infty$, we have $\mathcal Q_{2r}\longrightarrow w_1+w_{-1}$ and $
\mathcal Q_{2r+1}\longrightarrow w_1-w_{-1}$.

These statements follow from the argument of
Theorem~\ref{thm:testfunctions} in either branch: the first relevant
degree is $2k+4$ in the odd case and $2k+5$ in the even case.
The argument requires only nonnegative weights and $w_1>0$.
Indeed, for the converse, division
$P_j^{(n)}=D_jf_m+U_j$, with $\deg U_j<m$, gives the admissible
polynomial $G_c=P_j^{(n)}-U_j+cf_m$ for sufficiently large $c>0$,
and
\[
 G_c(1)-\frac{(G_c)_0}{w_1}=\frac{\mathcal Q_j}{w_1}<0.
\]
Without strict coefficient positivity, a negative test value alone
does not justify this perturbation. Moreover, checking finitely many
degrees does not certify nonnegativity for all $j\ge m+1$ unless
the remaining degrees are controlled by a separate argument (see~\cite[Sections 4.3--4.4]{BDHSS2016Energy} for such arguments).

\item \textbf{Investigate attaining codes.}
If an $(a,b)$-avoiding $s$-code of cardinality $N=1/w_1$, which must then
be an integer attains the bound, then the equality conditions in Theorem~\ref{LP-theorem}
require its inner products to be zeros of $f_m$ and $(f_m)_iM_i(C)=0$ for every $i\ge1$. Furthermore, if $(f_m)_i>0$ for
$1\le i\le m$, such a code is a spherical $m$-design. These necessary conditions guide the search for an extremal code. 
\end{enumerate}

\subsection*{Extension to energy bounds}

The same quadratures give a parallel programme for energy.
For either branch, verify the hypotheses of
Theorem~\ref{thm:energy} or~\ref{thm:evenEnergy}, including
positivity of the quadrature, the location of its nontrivial nodes
in $X_{a,b,s}$, and
\[
 N=\frac1{w_1}\in\mathbb Z_{\ge2}.
\]
Establish a positive Newton ordering of the interpolation multiset
\eqref{eq:interpolationMultiset} or
\eqref{eq:evenInterpolationMultiset}. The structural sufficient
conditions are \textup{(NO1)}--\textup{(NO4)} in the odd branch
and \textup{(ENO1)}--\textup{(ENO4)} in the even branch.
These include partial-product positivity and the full pairwise
Krein conditions, in addition to the conditions used for the
cardinality bounds. When $k=0$, explicitly require
$(t-a)(t-b)\in\Fpos$ or $(1+t)(t-a)(t-b)\in\Fpos$, respectively, as specified in
Remarks~\ref{rem:oddNewtonKzero} and~\ref{rem:evenNewtonKzero}.

For every absolutely monotone potential $h$ on $[-1,1)$, with
the stated endpoint convention at $-1$, the Hermite interpolants
then yield~\eqref{eq:ULB} and~\eqref{eq:evenULB}.
These bounds apply to all $(a,b)$-avoiding codes of cardinality
$N$, without imposing the upper bound $s$ on their inner products.
They also remain valid if only the derivatives of positive order
are nonnegative and $h$ itself takes negative values.

Finally, the same conditions $\mathcal Q_j\ge0$ for every $j\ge m+1$
certify optimality in the full polynomial energy linear programme.
For strictly absolutely monotone $h$, a negative test value gives
a strict improvement, so these conditions are also necessary.
The odd statement is Theorem~\ref{thm:testfunctionsULB}; the same
proof applies in the even branch with its quadrature, Hermite
interpolation multiset, and threshold $2k+5$.
This energy criterion does not require strict positivity of the
coefficients of $f_m$. Under the combined hypotheses for both
criteria, including strict coefficient positivity for cardinality
and strict absolute monotonicity for energy, the two bounds are
simultaneously LP-optimal or simultaneously improvable.

Thus, in both branches, exactness follows from truncated
orthogonality. The main tasks are to establish the required root
locations, the signs of the fixed-node weights, and the
appropriate Gegenbauer positivity conditions. The Hermite--Newton
construction additionally uses Newton positivity, while the
test-function criterion for global LP optimality involves all
higher degrees.

\end{document}